\documentclass[manuscript]{acmart}

\usepackage{pifont}
\usepackage{subcaption}
\usepackage{tabularx}
\usepackage{longtable}
\usepackage{array}
\usepackage{makecell}
\usepackage{multirow}
\usepackage{rotating}
\usepackage{geometry}
\usepackage{xcolor}
\usepackage{lscape}
\usepackage{natbib}

\AtBeginDocument{%
  }

\setcopyright{rightsretained}
\copyrightyear{2025}
\acmYear{2025}
\acmDOI{XXXXXXX.XXXXXXX}

\acmJournal{CSUR}

\begin{document}

\title{Deep Learning based Key Information Extraction from Business Documents: Systematic Literature Review}

\author{Alexander Michael Rombach}
\orcid{0000-0002-9173-4215}
\affiliation{
  \institution{Saarland University}
  \department{Institute for Information Systems}
  \city{Saarbrücken}
  \state{Saarland}
  \country{Germany}
}
\affiliation{
  \institution{German Research Center for Artificial Intelligence (DFKI)}
  \department{Institute for Information Systems}
  \city{Saarbrücken}
  \state{Saarland}
  \country{Germany}
}

\email{alexander_michael.rombach@uni-saarland.de}

\author{Peter Fettke}
\orcid{0000-0002-0624-4431}
\affiliation{
  \institution{Saarland University}
  \department{Institute for Information Systems}  
  \city{Saarbrücken}
  \state{Saarland}
  \country{Germany}
}
\affiliation{
  \institution{German Research Center for Artificial Intelligence (DFKI)}
  \department{Institute for Information Systems}  
  \city{Saarbrücken}
  \state{Saarland}
  \country{Germany}
}
\email{peter.fettke@dfki.de}


\begin{abstract}
    Extracting key information from documents represents a large portion of business workloads and therefore offers a high potential for efficiency improvements and process automation. With recent advances in Deep Learning, a plethora of Deep Learning based approaches for Key Information Extraction have been proposed under the umbrella term Document Understanding that enable the processing of complex business documents. The goal of this systematic literature review is an in-depth analysis of existing approaches in this domain and the identification of opportunities for further research. To this end, 130 approaches published between 2017 and 2024 are analyzed in this study.
\end{abstract}



\maketitle

\section{Introduction}
\label{intro}

The general idea of a paper-free -- or at least paperless -- office already came up five decades ago \citep{paperless}. However, to this day, physical paper documents still play an important role in business operations, as they are a key means of communication related to transactions both within and between organizations \citep{bd_benchmarks}. 
The processing of such documents is an essential yet time-consuming task that offers a high potential for automation due to the high workload involved as well as the critical nature of information transfer between different information systems \citep{R3, ie_critical}.
At the same time, it can be observed that the ongoing digital transformation of business operations is leading to an increase in the digital processing of documents. This trend reinforces the need -- but also the potential -- for automated document processing, as more and more documents are available in digital form \citep{data_requirements}. 

Research on document processing is not new and has been conducted for several decades \citep{DENGEL2004}. In fact, the term ''document analysis'' can be traced back to the 1980s \citep{1980s}. In recent years, however, there has been an upsurge in research related to document processing based on visually-rich documents (VRDs) and business documents, made possible by major advances in Deep Learning (DL). 
One of the most studied document processing task in this regard is Key Information Extraction (KIE) \citep{R6}, which is concerned with extracting specific named entities from documents in a structured form \citep{bd_benchmarks}.

Complex business documents pose a significant challenge to KIE systems because they cannot be understood and processed as linear text sequences, as has been the case in most traditional KIE applications \citep{DUE}. These documents typically contain implicit and explicit cues and complex positional dependencies between certain text segments. In this regard, related research investigates different methods to integrate such cues into the model architectures in order to achieve better extraction results.
Business documents also have special characteristics resulting from their connection to business processes. For example, different documents that are being processed as part of the same process run usually contain reoccurring information. These aspects and how they can improve corresponding KIE systems are worth investigating. In general, a process-oriented understanding is critical in order to adequately address the challenge of KIE in real-world contexts.

Although many DL-based KIE approaches for VRDs have been proposed, there is no comprehensive overview of most recent work in this area that focuses on the underlying DL concepts and technical characteristics while also adopting a business process perspective. The aim of this systematic literature review (SLR) is to fill this gap and provide a detailed overview of this research area and its state of the art. The contribution of this work is threefold:

\begin{enumerate}
    \item A SLR based on 130 approaches to provide a concise overview of DL-based KIE methods for business documents.
    \item Categorization and detailed comparison of corresponding methods based on various characteristics.
    \item Dissemination of results, research gaps and derived potentials for future research.
\end{enumerate}

The remainder of this manuscript is structured as follows:
Chapter two provides background information on key concepts and nomenclature that are relevant to this SLR. 
In chapter three, we discuss related work in terms of existing surveys and illustrate, how this study differs.
The methodology used for this SLR, and more specifically how relevant literature was identified, is illustrated in chapter four.
Chapter five provides an overview of the results of the in-depth analysis.
Chapter six serves as a discussion of the key findings and is dedicated to specific aspects of the identified literature.
Based on the previous two chapters, we propose a research agenda and starting points for follow-up research in chapter seven.
Chapter eight concludes the manuscript with a summary.
\section{Background}
\label{background}

\subsection{A business process perspective}
\label{kie_processes}

Although business documents have a value of their own, they are best understood in relation to each other. Whenever they belong to the same process run, the information contained in these documents is also closely related to each other. For example, all documents related to the same run of a purchasing process typically contain the same order ID.
Because they are embedded in specific processes that specify what information is relevant, business documents always contain distinct sets of predefined entities \citep{future_paradigms_bd}. For example, invoices contain many different types of monetary values and unique identifiers such as invoice numbers that need to be identified. 

Understanding temporal relationships in business processes is also important. Consider the following example of a simplified purchasing process. First, a customer places an order for a physical good. The company then processes the order, which results in the company sending an order confirmation and, at a later stage, the actual physical goods, a delivery note and an invoice. Figure SF1 of the electronic supplementary material shows an exemplary run of a corresponding process, including the interactions with internal and external entities such as customers and suppliers. Sequences where business documents are relevant are highlighted in red, indicating parts of the process where Document Understanding approaches could play a key role. As can be seen, especially interfaces related to external partners are expressed through document-based exchanges.

In addition to implicit knowledge about predefined entities, the consideration of associated business processes can also be useful in the sense that the information flow between individual process steps could be taken into account. For example, one could consider how the data extracted from a business document is further processed in subsequent process steps. From this, conclusions could be drawn as to how corresponding business documents should be processed. 
One could also analyze whether -- in addition to the documents themselves -- there is an additional data flow between the document processing steps that could facilitate the understanding of the documents.

\subsection{Document Understanding}

Document Understanding (DU), based on concepts from Natural Language Processing (NLP) and Computer Vision (CV), is an umbrella term covering a wide range of document processing tasks, including KIE, table understanding, document layout analysis (DLA), document classification and visual question answering (VQA) \citep{DUE}.
More recently, novel tasks have been proposed such as ''Key Information Localization and Extraction'', which is an extension of KIE that emphasizes the need to localize key information, e.g., by identifying bounding boxes\footnote{A bounding box indicates the coordinates of an element in the document image by its top left corner as well as its width and height.} \citep{bd_benchmarks}.
For a brief introduction to DU tasks besides KIE that are not covered in this review, we refer to \cite{R3}.
There are also other terms that are used for this research area, such as Document Analysis, Document AI or Document Intelligence \citep{docintelligence, R3}. 

DU can also be grouped according to the complexity of the tasks, as suggested by \cite{nr55}: Perception tasks deal with the recognition of descriptive document elements (e.g., text or entire tables), whereas induction tasks aim at the extraction of enriched information (e.g., document class or named entities) based on the perceived documents. Finally, reasoning represents the most complex subtask, combining perception and induction to enable document understanding beyond the explicitly contained information, mostly in the context of VQA. An example of a reasoning task is obtaining natural language explanations of figures and diagrams in documents \citep{bd_benchmarks}.

\subsection{Key Information Extraction}
\label{kie}

\begin{figure}[!ht]
    \centering
    \includegraphics[width=0.5\textwidth]{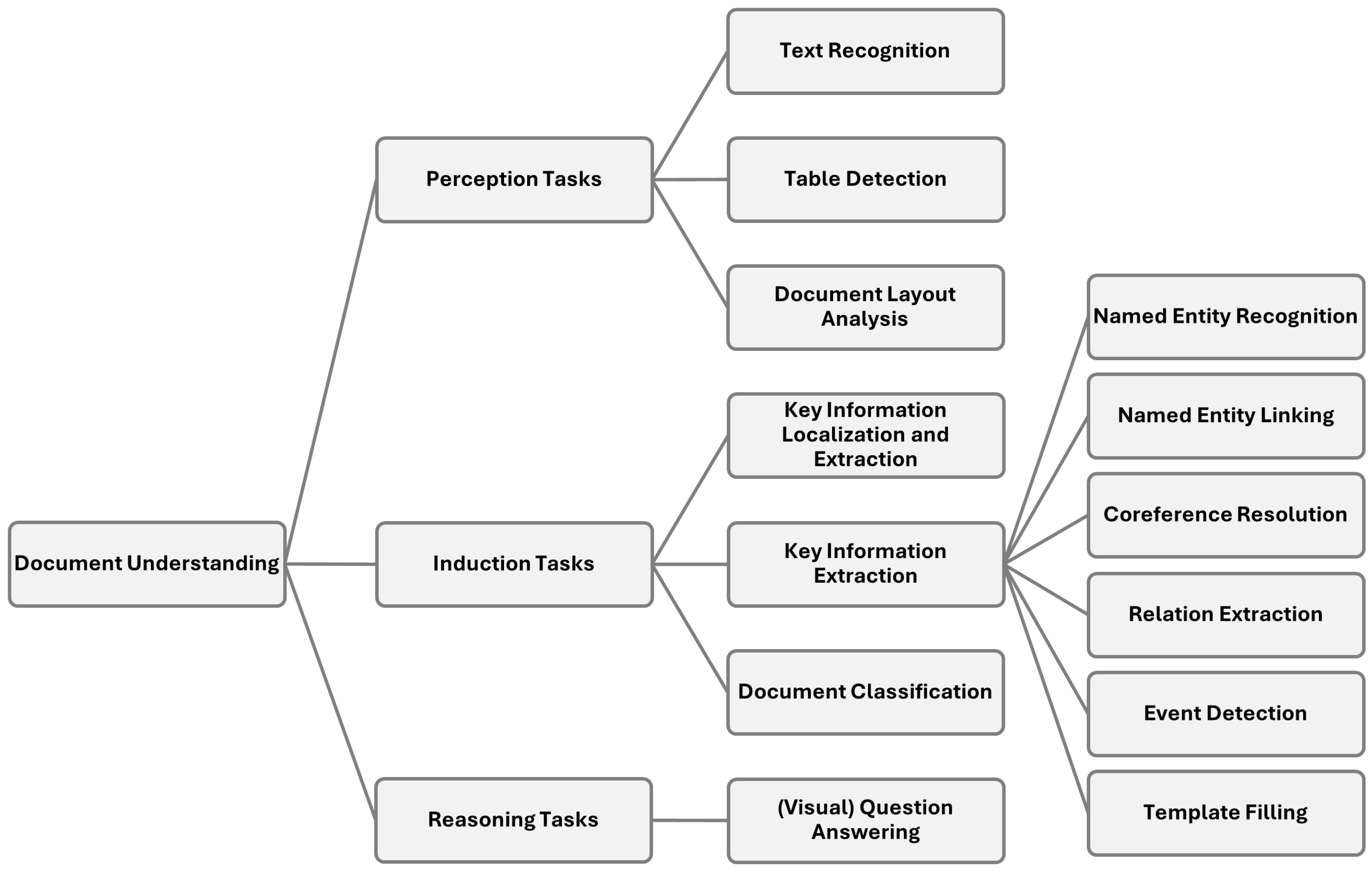}
    \Description{A hierarchy of tasks and subtasks related to Document Understanding and Key Information Extraction}
    \caption{Overview of Document Understanding and Key Information Extraction}
    \label{fb}
\end{figure}

KIE investigates methods for the automated extraction of named entities from documents into structured formats \citep{bd_benchmarks} and can be further subdivided into individual research areas, namely Named Entity Recognition (NER), Named Entity Linking, Coreference Resolution, Relation Extraction (RE), Event Detection and Template Filling \citep{kie_1, nr17}, with NER and RE being the most prominent ones. The goal of NER is to detect entities in text and assign predefined labels to them, usually solved as a sequence labeling task \citep{Balog2018}.
Figure \ref{fb} visualizes the outlined hierarchy of DU and subordinate research areas such as KIE.\footnote{Note that this hierarchy is not a result of this study, but rather a reflection of the understanding in the literature.}
    
KIE approaches can be divided into three main subgroups based on how they represent the underlying documents, namely graph-based, grid-based and sequence-based methods.
Graph-based systems convert document pages into graph structures that represent the layout and content of the pages. Such graphs typically allow for a flexible structure and can be constructed in a variety of ways. For example, each word or even character in the document image can be considered as a node in the graph. Different setups are also possible regarding the definition of the edges, for example creating a fully connected graph where every node is connected to every other node.
Instead of constructing graphs, the grid-based approaches aim to create more organized grid structures with well-defined connections primarily characterized by rectilinear links -- often based on the document image pixels. The grids are usually defined on different granularity levels, which affects which features the individual grid elements will be assigned to. For example, if the grid is defined on character level, all grid elements that are overlapped by the bounding box of a given character in the document image will have a value that is derived from that character (e.g., constant character index).
Sequence-based techniques, on the other hand, convert documents into linear input sequences, ideally preserving and incorporating the document layout and other visual cues. These input sequences are then usually processed by Sequence Labeling methods, where each element is assigned to a particular class. For more details, we refer to \cite{R8}.
The following figure \ref{document_representations} illustrates the aforementioned paradigms and how they represent a sample document snippet. 

\begin{figure}[!ht]
    \centering
    \begin{subfigure}[b]{0.33\textwidth}
        \centering
        \includegraphics[width=\linewidth]{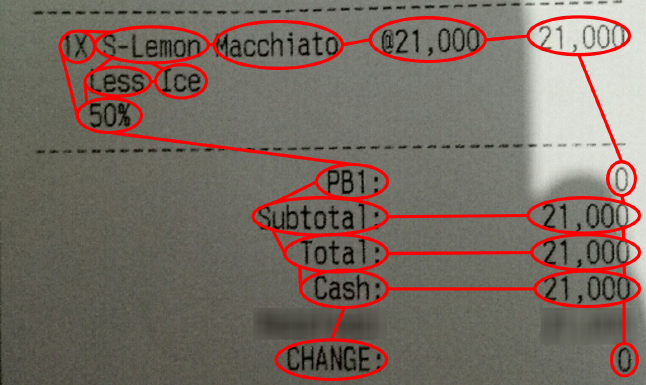}
        \caption{Graph-based (word-level)}
        \label{example_graph}
    \end{subfigure}
    \hspace{0.1\textwidth}
    \begin{subfigure}[b]{0.33\textwidth}
        \centering
        \includegraphics[width=\linewidth]{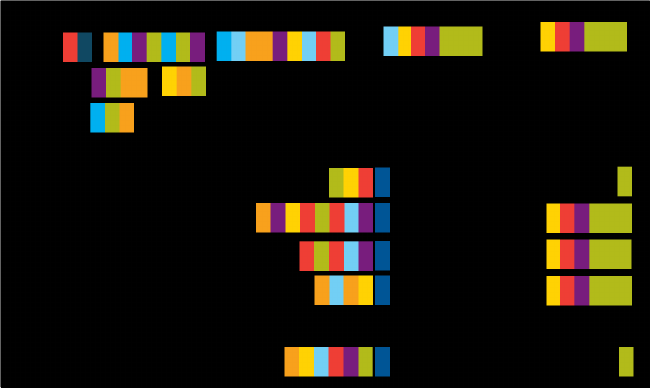}
        \caption{Grid-based (character-level)}
        \label{example_grid}
    \end{subfigure}
        
    \begin{subfigure}[b]{0.8\textwidth}
        \centering
        \includegraphics[width=\linewidth]{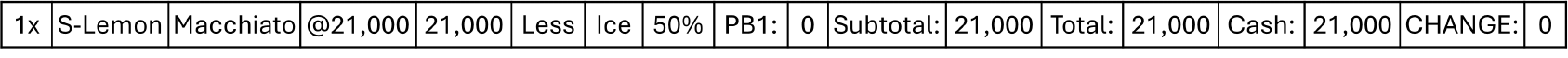}
        \caption{Sequence-based (word-level)}
        \label{example_sequence}
    \end{subfigure}
    \Description{Different representations of a document snippet based on graph-based, grid-based and sequence based structures with different granularity levels}
    \caption{Main document representations of KIE methods (best viewed in color)}
    \label{document_representations}
\end{figure}

\subsection{Input documents}
\label{terminology}

In the context of DU research, input documents are often referred to as "visually-rich documents". Although there is no universally accepted definition, an aggregated understanding can be derived as follows: VRDs typically represent complex documents in which the simultaneous consideration of text, layout and visual information is of great importance in order to adequately capture their semantics \citep{R3, nr49, nr2, nr13}.
Examples of visual features are font properties such as bold text segments with an increased font size that represent a document title or specific keywords that indicate information to be extracted \citep{R3}. Thus, all modalities are important for KIE and converting VRDs to linear text sequences would otherwise result in a significant loss of information. This is in contrast to "simpler" documents such as news articles, where representations as linear text sequences are usually sufficient. This is also the reason why advances in DL enabled the processing of such complex input documents, as they allow for the integration and adequate processing of different input modalities.

The term "business document" appears less frequently in related literature and there is also no universal definition of such documents \citep{future_paradigms_bd}. In general, business documents contain process-relevant details related to the internal and external operations of an organization, as these documents represent a central means of communication \citep{R3, bd_benchmarks}. 
Business documents, like VRDs, pose a significant challenge to automated DU systems for a variety of reasons, as discussed by \cite{docintelligence}. For example, corresponding documents exist in a variety of different formats and are often only available in scanned form due to their paper-based distribution. A scanned document is reduced to its image data and therefore requires techniques such as Optical Character Recognition (OCR) to make the content machine-readable \cite{scan_vs_native}. The layout and overall content of business documents can vary from highly structured to highly unstructured. Furthermore, business documents can have relationships to other documents and/or may consist of multiple documents in a hierarchical fashion. To understand business documents, it is therefore necessary to observe such interrelationships and to understand the temporal relationships in processes.

As mentioned at the beginning, document processing is a central activity in business contexts. Therefore, it is promising to study KIE from a business process perspective and to investigate methods that directly address this challenge.
The scope of this work regarding input documents is the intersection of VRDs and business documents. Therefore, we do not consider VRDs that are not typically embedded in business contexts, and at the same time, we do not consider business documents without characteristics of VRDs. An example of the latter is a company's general terms and conditions. Although it can be considered as a business document, it typically appears as a simple text document without enriched visual elements.
We use the terms VRDs and business documents interchangeably when addressing this intersection.

\section{Existing reviews on Key Information Extraction}
\label{related}

Many surveys are either domain-specific (e.g., healthcare \citep{clinical_slr}) or cover approaches that deal only with linear text inputs \citep{text_slr}. In recent years, a few surveys have been published that cover DL-based DU methods while also dealing with VRDs. However, not all of them focus on the task of KIE (e.g., \cite{dla_slr} cover DLA). 

In the following, we discuss the most closely related work.
\cite{R1} provide a brief KIE section that is limited to discussions on a few relevant aspects. The survey also does not include an in-depth analysis of the approaches.
In \citep{R2}, the authors provide a comprehensive overview of document processing with many different facets. However, it offers limited discussion about KIE. Much refers specifically to NER, where the overall workflow as well as different methods regarding pre-processing and feature extraction are being discussed. The authors discuss some model architectures. Since their chosen search period includes work between 2010 and 2020, a large number of analyzed papers also do not necessarily use DL methods and/or consider VRDs.
\cite{R3} cover a wide range of DL-based DU topics. KIE, on the other hand, is discussed relatively briefly. The survey discusses aspects related to transformer-based document processing methods from a general point of view. However, details about the employed model architectures and other KIE paradigms are not covered in detail.
The survey of \cite{R4} focuses on preliminary DU tasks such as text detection and text recognition. KIE is also covered, although only a small number of approaches are included. The authors present some individual approaches in detail, however there is a lack of comprehensive overview.
In \citep{R6}, document processing is positioned in the context of Robotic Process Automation. Although an extensive literature survey is conducted, there is no in-depth technical discussion of KIE approaches for VRDs.
An overview of key methods for KIE is provided by \cite{R7}, albeit at a relatively high level. The authors however provide an extensive outlook on future work.
Besides KIE, \cite{R8} also cover VQA and document classification. The authors propose a taxonomy for DU along different dimensions and focus on KIE benchmark datasets. Some challenges as well as future work are also discussed.
\cite{R9} present a review of transformer-based methods for DU tasks. They highlight key paradigms of corresponding models and showcase a few approaches in detail. A major focus of the review is the description of benchmark datasets and related performance comparisons. In general, the survey chooses to discuss a few transformer-based approaches in detail, but does not present a broad overview of DL-based approaches with respect to KIE.

Overall, existing surveys either examine KIE from a very broad perspective, cover only a few approaches in detail, or alternatively include a higher number of published work, but at the expense of a detailed discussion of the underlying methods. 
This work on the other hand provides an in-depth analysis, both quantitatively and qualitatively. 
One key distinguishing factor is also the adoption of a business process perspective. As discussed in section \ref{kie_processes}, the consideration of business processes as well as general domain knowledge is of high importance for KIE systems. To this end, only the review by \cite{R6} also adopts a practice-oriented perspective during the analysis.
Table \ref{slrs} summarizes the differentiation of the previously mentioned related work along certain distinguishing factors. 

\begin{table}[!ht]
    \caption{Comparison against existing reviews}
    \label{slrs}
    \begin{tabularx}{\textwidth}{l *{9}{>{\centering\arraybackslash}X}}
    \toprule
    \textbf{Distinguishing factor} & \textbf{\cite{R1}} & \textbf{\cite{R2}} & \textbf{\cite{R3}} & \textbf{\cite{R4}} & \textbf{\cite{R6}} & \textbf{\cite{R7}} & \textbf{\cite{R8}} & \textbf{\cite{R9}} & \textbf{Ours} \\ 
    \midrule
    Search period & n/a & 2010-2020 & n/a & n/a & 2017-2022 & n/a & n/a & 2014-2023 & 2017-2024 \\ 
    Detailed search strategy &  & ***** & \ & \ & ***** & \ & \ & **** & ***** \\
    Focus on KIE task & ** & *** & *** & *** & ** & ***** & *** & *** & ***** \\
    Business process perspective & \ & ** & * & \ & **** & \ & \ & \ & ***** \\
    Analysis of DL concepts & * & *** & ** & ** & * & ** & ** & **** & ***** \\
    Analysis of key characteristics & ** & *** & **** & *** & *** & **** & **** & *** & ***** \\
    Analysis of KIE datasets & * & * & *** & * & * & **** & *** & ***** & ***** \\
    Performance comparison & \ & ** & \ & ** & * & ** & \ & **** & ***** \\
    Consideration of future work & \ & *** & ** & * & ** & *** & *** & * & ***** \\
    \bottomrule
    \end{tabularx}
\end{table}

\section{Methodology}
\label{method}

\subsection{Research questions}

The overall research question that is to be answered by this SLR is: \textit{"What is the state of the art in Deep Learning based Key Information Extraction from business documents?"}.
To this end, eight further research questions have been defined:

\begin{itemize}
    \item RQ1: Which input modalities are considered and how are they integrated?
    \item RQ2: Which DL architectures are being used?
    \item RQ3: According to which criteria can existing approaches be categorized?
    \item RQ4: Which input documents are considered?
    \item RQ5: To what extent are practical applications and domain knowledge discussed?
    \item RQ6: Which are the best performing approaches?
    \item RQ7: Are there noticeable trends in the proposed approaches and architectures?
    \item RQ8: What potential for improvement can be formulated for follow-up research?
\end{itemize}

The first five research questions cover an in-depth analysis of the identified approaches, with the aim of examining the proposed methods and how they differ from each other. This includes key aspects such as input modalities, model architectures and data bases.
Based on this, research questions six to eight adopt a more aggregated view and aim to identify the state of the art and derive recommendations for follow-up research.

\subsection{Search procedure}

The following six databases were used to obtain relevant literature: ACM Digital Library (ACM), ACL Anthology (ACL), AIS eLibrary (AIS), IEEE Xplore (IEEE), ScienceDirect (SD) and SpringerLink (SL).
The search strings were carefully designed to include the commonly used terms for the research area as well as relevant keywords for the target domain (VRDs) and the application of DL architectures. 
If supported by the search engine, we also included terms that should not appear to filter out papers that are outside the scope of this work.
The search period was limited to the range 2017 to 2024. Considering 2017 as the lower limit is a significant help to avoid false positive results, as KIE related literature typically did not use DL methods before 2017.
The final search strings for each database, including the number of results at the time the query was run, can be found in table ST1 of the electronic supplementary material.

We defined the following inclusion and exclusion criteria as the basis for the literature screening. First, the title, abstract and conclusion were analyzed. If no violation of the criteria was identified on the basis of these parts, the full texts were analyzed subsequently.

\begin{itemize}
    \item IN1: The work is peer-reviewed and related to the research area of KIE.
    \item IN2: The work employs DL concepts and outlines its architecture.
    \item IN3: The work evaluates the effectiveness of the approach in a quantitative and/or qualitative manner.
    \item EX1: The work is not written in English and/or was published outside of the defined search period.
    \item EX2: The work does not propose a new approach for KIE, but rather applies existing approaches to different use cases, conducts a survey of existing approaches and/or only proposes a novel KIE dataset.
    \item EX3: The work does not apply the approach to VRDs, but to other domains and/or considers text-only inputs.
    \item EX4: The work does not focus on KIE, but on other DU tasks and/or focuses only on image pre-processing tasks.
    \item EX5: The work focuses only on extracting information from very specific document elements such as tables.
\end{itemize}

The overall search process according to the PRISMA \cite{prisma} is visualized in figure SF2 of the electronic supplementary material. A total of 130 relevant approaches were identified for this SLR. A list of the identified approaches as well as their numbering, which is also used as a reference in this manuscript, can be found in table ST2 of the electronic supplementary material.

\section{Results}
\label{results}

\subsection{Overview}
\label{overview}

\begin{figure}[!ht]
    \centering
    \begin{subfigure}{.33\textwidth}
        \includegraphics[width=\linewidth]{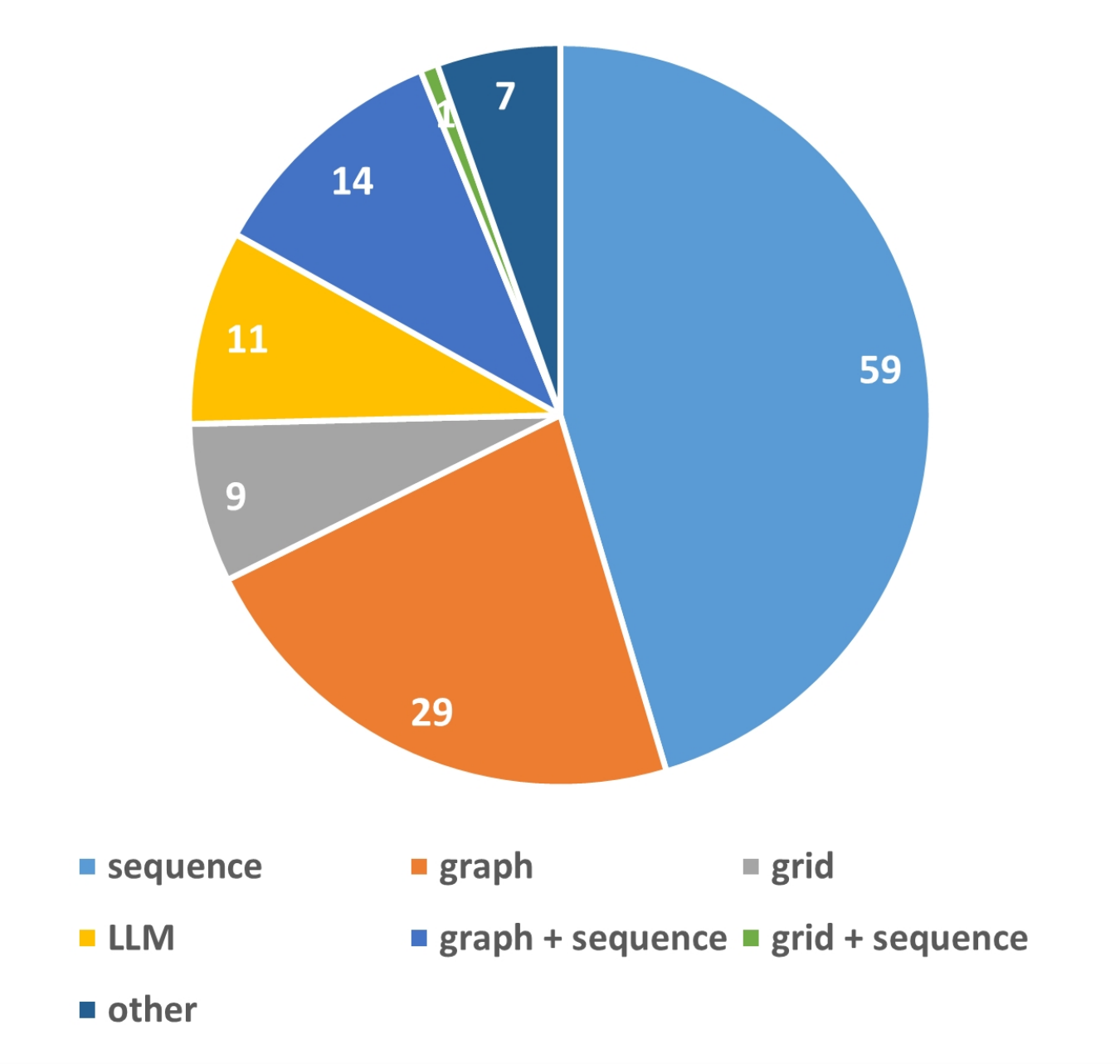}
        \caption{Overall distribution}
        \label{distr_overall}
    \end{subfigure}
    \hfill
    \begin{subfigure}{.66\textwidth}
        \includegraphics[width=\linewidth]{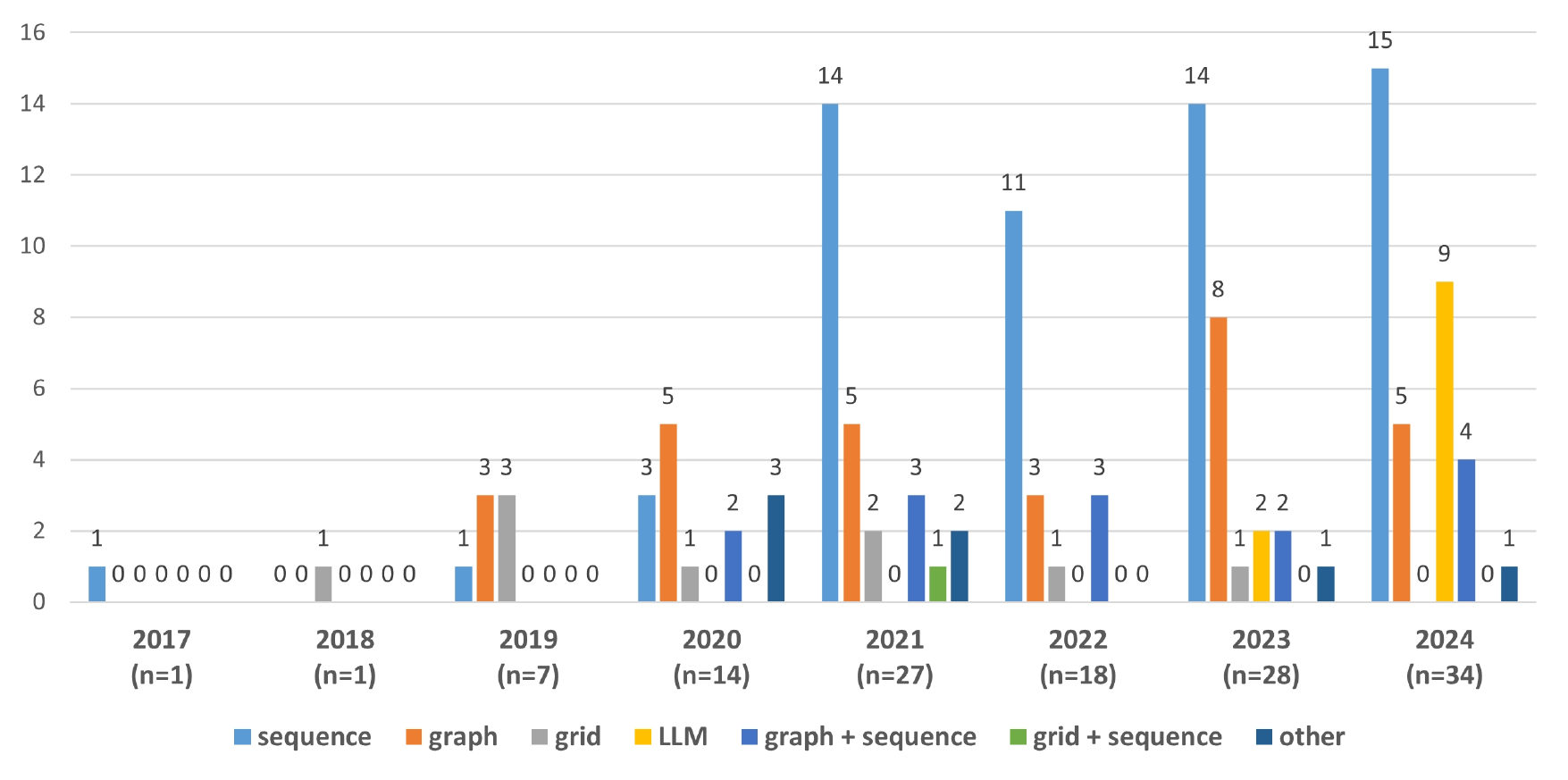}
        \caption{Distribution over time}
        \label{distr_time}
    \end{subfigure}
    \caption{Distribution of categories}
    \Description{A pie chart indicating the distribution of categories and a bar chart indicating the occurrences over time}
    \label{distr_categories}
\end{figure}

First, we provide a high-level overview of the analyzed work. To this end, table ST3 of the electronic supplementary material shows the results of analysis along different properties, each of which considers different aspects, also with practical applications in mind.
Figure \ref{distr_overall} visualizes the overall distribution of the KIE paradigms. Almost half of the approaches belong to the group of sequence-based methods. Two further common categories are graph-based systems and approaches that combine graph-based and sequence-based concepts. The grid representation is not as widely used as the graph representation, as only nine approaches are purely based on this paradigm. Large Language Model (LLM) based systems represent an emerging category with 11 identified approaches.
Also, only seven of the analyzed papers cannot be assigned to any of the groups, which underlines the dominance of the previously mentioned paradigms for KIE systems. Interestingly, only the work by \cite{nr48} combines grid-based and sequence-based concepts into one KIE approach.
Based on the distribution of the methods over time, as visualized in figure \ref{distr_time}, one can see an increased interest in KIE research, especially since 2021. It is also noticeable that the dominance of sequence-based approaches first started in 2021, while graph-based methods were the most common group of approaches before that. The increased popularity of sequence-based methods in this period can probably be attributed to the influential work by \cite{nr21}, which was among the first to show state-of-the-art results using a transformer architecture and extensive pre-training procedures, which in turn led to a lot of follow-up research in this direction.
There is also a large increase in graph-based methods in 2023 compared to previous years and compared to the other categories. This category has therefore remained relatively popular over time.
Overall, 2024 had the highest number of papers published, showing that there is still a high level of interest in this area of research. In 2022, however, there was a considerable decrease in the number of corresponding KIE approaches. One reason for this observation may be the aforementioned increased interest in sequence-based methods, which typically require extensive pre-training. Therefore, many scholars spent 2022 on developing corresponding methods and on academic write-up, leading to a publication some time later in 2023 or 2024. Given the emergence of LLMs, they also start to appear in KIE research in 2023, with a significant increase in 2024, already becoming the second most common category.

The analysis of the integrated input modalities shows that the vast majority of approaches integrate at least textual and layout-oriented features. This is not surprising, since these elements contain the most relevant information for document processing. Layout modalities are usually derived from bounding boxes, i.e., the coordinates of words in document images.
Around 60\% of the analyzed approaches integrate visual features obtained from the document images. As discussed in section \ref{terminology}, the integration of such visual cues into KIE systems can be crucial for a proper understanding of complex VRDs. It could be argued that -- given this relevance -- there are relatively few approaches that integrate image modalities. Nevertheless, the results show that visual cues have been increasingly integrated into the models over time, especially since 2021. In this regard, \cite{nr62} extensively investigate different methods to obtain and fuse visual modalities into KIE systems. They show that the choice of how visual information is represented depends on the underlying VRDs and that attention-based fusion mechanisms can outperform more basic methods such as concatenations of features vectors.
In general, integrating multiple modalities can play an important role for DU systems, as they incorporate cues from different aspects that complement each other. For example, the image modality can provide relevant insights for documents with otherwise limited amounts of text and bounding boxes, while the layout modality is crucial for documents with very complex layouts and distinct structures \citep{aesthetics}.
Some approaches use hand-crafted input modalities. In most of these cases, they are integrated through custom input features such as boolean flags, whether a word in the document represents specific entities such as dates or monetary values. Another feature type is information about the position of a particular word with respect to the overall reading order. In some papers, hand-crafted inputs are based on syntactic features such as the numbers of characters of a word, details regarding the fonts or encoded character representations. 
The use of hand-crafted features is also often associated with grid-based approaches. Such methods usually define a custom encoding function that assigns a specific value to each element of the grid. For example, \cite{nr1} assign a constant value to each pixel in a document image, such as an integer index of a character at each position -- or the value 0 for blank parts of the document image. 
None of the approaches integrate meta data resulting from related business workflows as discussed in section \ref{kie_processes}, therefore lacking a practice-oriented perspective in this regard.
Overall, hand-crafted features provide additional information beyond data that is explicitly contained in documents and can thus improve KIE systems. It must be said, however, that hand-crafted input features can potentially lead to a considerable amount of additional work in terms of labor-intensive data annotation \citep{data_requirements}, which could also explain the less frequent use of this feature type. One could also argue that the inclusion of such hand-crafted features can potentially result in overfitting to the domain of the training documents -- leading to poorer generalization capabilities across different domains. However, there is no work that analyzes this trade-off between integrating hand-crafted input features and the generalization capabilities across different domains.

In terms of the underlying data basis, the median number of employed documents is 2,172. This is a relatively low amount, especially since the data basis usually has to be split into training and evaluation partitions. As a result, many of the approaches are often only trained on a small document corpus and thus probably with little variety in terms of layouts. 
Since DL-based DU models usually require large amounts of data for training \citep{data_requirements}, it is debatable whether the proposed approaches have reached their full potential. Note that these observations do not include the data basis that is being used for pre-training purposes, which will be discussed in the context of sequence-based approaches in section \ref{sequence}.
Deviating from the average, \cite{nr8,nr34,nr44} employ more than one million documents for implementing their KIE system. On the other hand, in some cases, only 199 documents are used for developing the models, while still achieving promising extraction results. This indicates that there are major differences in terms of data requirements of the individual approaches and that some architectures can manage with significantly less data.
The vast majority of considered document types are receipts and forms, which are used by at least half of the approaches. The main reason for this is the fact that the most frequently used benchmark datasets are based on these document types. Besides, invoices are most often considered when authors do not use public benchmarks, but rather private in-house datasets.  

Regarding the aspect of reproducibility, implemented code is available for 50 approaches. In most cases, authors share their code directly, however there are exceptions where implementations have been made available by external sources through re-implementation. Model weights are even less commonly shared, although this would be particularly helpful for pre-trained models, as improvements and optimizations could be made by external parties on this basis. When either code or model weights are shared, they are mostly provided with a license that allows for commercial use. This is beneficial for organizations that want to integrate corresponding KIE systems into their own business processes, which in turn helps to disseminate the research efforts.

A few approaches are independent of external OCR engines and are therefore either responsible both for text reading and information extraction (end-to-end) or alternatively require no text reading stage at all and map input VRDs immediately to desired outputs (OCR-free) \citep{R8}. The system by \cite{nr24} is only end-to-end during the pre-training phase. Also, \cite{nr34} propose a system that can be designed in an end-to-end manner, however the authors chose an external OCR engine in their work. These approaches first appeared in 2019, but became more frequent since 2022.
The same is true for the 23 identified generative KIE methods, which first appeared in 2021, but now see an increased interest, especially given the rise of generative LLMs. This category of KIE methods is able to output arbitrary texts using autoregressive decoding mechanisms. Key advantages are, that corresponding systems are not necessarily influenced by faulty OCR extractions as they can generate OCR-free representations of the target words \citep{nr53}. They can also be more easily adapted to different DU tasks, for example by using distinct textual prompts \citep{nr59,nr72}. More traditional KIE methods that use a classification head would on the other hand need to be retrained with a different classification layer in order to perform a different task like document classification \citep{nr72}.

Only 10 papers explicitly use domain knowledge. The integration of domain knowledge mostly consists of hand-crafted input features, as discussed before. This again emphasizes the lack of a practical perspective in related literature, as corresponding KIE systems do not consider insights that can be obtained from real-world workflows. 
Also, only three approaches have been evaluated in real-world industry settings. In these cases, the authors show the impact of the developed KIE methods on real-world document processing tasks -- for example in terms of efficiency improvements. Therefore, there is a severe gap between research advancements and real-world applications as the majority of proposed models are not evaluated in practical settings. It is therefore possible that innovations that improve benchmarks may not be immediately suitable for real-world scenarios, e.g., due to extensive data requirements.
16 of the analyzed papers represent an evolution of an existing approach. A prominent example is the LayoutLM family. Based on the original LayoutLM \citep{nr21}, many subsequent refinements and improvements in various aspects have been proposed.
Another example is Chargrid \citep{nr2}, which has been the basis for many other grid-based approaches (see section \ref{grid}). 
A few of the analyzed approaches are implemented using weakly-annotated data. In most cases, this means that besides the document images, only the textual target values are required. Some other approaches only require annotations on segment-level or word-level. Corresponding approaches therefore do not need to be trained on fully annotated documents, which usually involves a labeling of each word including its bounding box and entity. Such approaches can be more suitable for real-world applications where data annotation is costly.

\begin{figure}[!ht]
    \centering
    \includegraphics[width=0.9\textwidth]{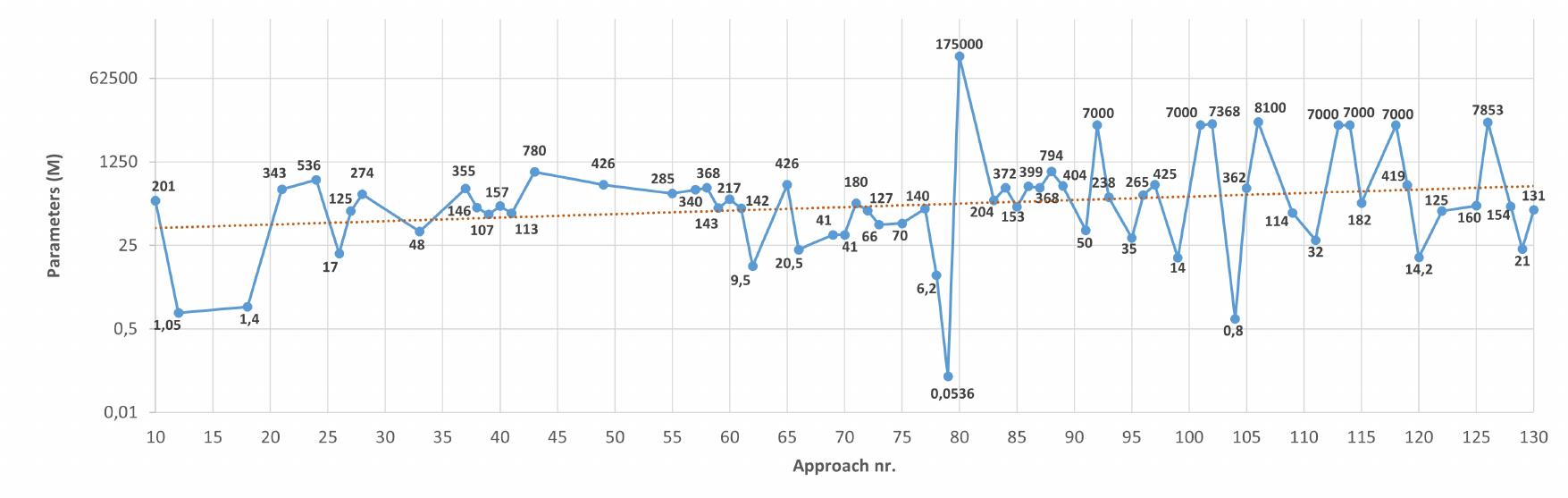}
    \caption{Evolution of parameter counts}
    \Description{A chart that shows the evolution of parameter counts. The graph shows that there was no steep increase over time.}
    \label{params}
\end{figure}

The evolution of model sizes in terms of their number of parameters (in millions) is shown in figure \ref{params}. Note the necessary logarithmic scale of the y-axis due to the approach by \cite{nr81}, which is based on GPT3 and therefore includes 175 billion parameters. The orange line indicates the trend over time. Based on the analyzed approaches, the median number of trainable parameters is 170 million. Most of the approaches tend to have between 20 and 500 million parameters, which did not change significantly over time. However, with the advent of LLMs, more and more approaches with several billion parameters have been proposed, as shown in the figure.

\subsection{Categories}
\subsubsection{Sequence-based methods}
\label{sequence}

Table ST4 of the electronic supplementary material presents the results regarding the analysis of sequence-based approaches. We also include hybrid KIE approaches that are based on multiple paradigms (e.g., a combination of graph-based and sequence-based methods) in this analysis and present the findings regarding the sequence-based subsystem.
Most of the methods are defined on word-level, i.e., each word of a document image represents one element in the sequence. In some cases, a segment or sentence-level granularity is chosen, which considers sentences as semantic entities and subsequently calculates aggregated embeddings. Other possible granularities -- albeit rarely used -- are character-level, token-level or cell-level.
The word-level definition is popular because it balances semantic richness with computational efficiency. It also aligns well with the OCR outputs and pretrained transformers like LayoutLM. While more fine-grained definitions can improve structural understanding, they are less common due to increased complexity and alignment challenges. Some authors use multiple granularity types simultaneously, allowing the system to consider both granular and coarse features, which can complement each other.

Around half of the sequence-based approaches do not consider a pre-training of the models and subsequent fine-tuning steps for the KIE task. This is somewhat surprising, as it has been shown that obtaining a general document understanding through pre-training is a promising research direction.
If pre-training is conducted, the average number of documents used is around five million. This highlights the relatively large amount of data that is required to pre-train corresponding models. The largest document corpus for pre-training is used by \cite{nr52}, which consists of 43 million documents. Deviating from this is the approach of \cite{nr41}, which only uses 5,170 documents for pre-training procedures, while still achieving competitive benchmark results.
Pre-training is usually very resource-intensive. For example, \cite{nr21} report that their largest model variant required 170 hours to finish one training epoch of a dataset consisting of six million documents.
In around half of the cases, the IIT-CDIP dataset is used. This dataset has been proposed in \citep{iit_cdip} and includes documents of the Legacy Tobacco Document Library\footnote{\url{https://www.industrydocuments.ucsf.edu/tobacco/}} regarding lawsuits against the tobacco industry in the 1990s. Another common dataset is RVL-CDIP, which however represents a subset of IIT-CDIP and therefore includes no additional documents.
Also, in seven cases, a private document collection has been used for pre-training procedures. Besides these common datasets, there exist other large datasets such as DocBank, which are however rarely used.
Table ST4 also lists the employed pre-training tasks, with the most common being Masked Visual-Language Modeling (MVLM) proposed by \cite{nr21}. This task is strongly related to the more general Masked Language Modeling, which was originally introduced with BERT \citep{bert}. For the sake of MVLM, random tokens of the input sequences are masked where the goal of the model is to predicted the masked tokens given the surrounding context. Importantly, positional information is not masked and thus helps the model with the reconstruction. In this way, the model learns language contexts while also exploiting positional information. 
Some authors transfer this this idea to the image modality and employ Masked Image Modeling, where the model needs to reconstruct masked image tokens. However, a limitation of this task might be the fact that document images for the large part consist of white background and thus the model tends to simply predict white pixels without considering surrounding text and image tokens \cite{nr119}.
Various other related tasks have been proposed over time, which also explicitly include image modalities. This allows KIE models to fully exploit textual, positional and visual information during pre-training. 
When authors propose end-to-end and/or generative KIE methods, specific pre-training objectives are defined in order to learn text reading and producing outputs in an autoregressive manner. See also chapter \ref{train_setups} for a further discussion.

Regarding the distribution of integrated encoder and decoder architectures, 12 approaches do not incorporate a dedicated textual encoder at all, while 25 do not use a visual encoder. In \citep{nr1,nr81}, neither types of encoders are employed. 
In most cases, BERT or derived variants such as RoBERTA \citep{roberta} and SBERT \citep{sbert} are used for textual encoding. This is not surprising, given the popularity of BERT-based models in NLP. Another common encoding method is the LayoutLM-family, more specifically LayoutXLM, LayoutLM and LayoutLMv2. This can be explained by the popularity of these models in the context of DU as well as their extensive pre-training, which allows approaches to utilize expressive embeddings. This can also explain why many sequence-based approaches do not employ a dedicated pre-training stage, as they can make use of already pre-trained models. Some authors integrate more traditional embedding methods such as N-gram-models, Word2Vec or FastText instead.
Regarding visual encoders, one can see that there is no clear dominance of one particular model. Nonetheless, the most common visual backbone is represented by ResNet models \citep{resnet}, which are also popular in related CV tasks. ResNet50 and ResNet18 are the most often used variants of this architecture, however other variants that have been proposed as improvements such as ResNeXt101-FPN and ConvNeXt-FPN are employed as well.
Also, some authors incorporate the more recent Swin Transformer \citep{swin} as the visual encoder, which captures local and global image features through sequential processing of non-overlapping image patches and shifted window self-attention mechanisms.
Interestingly, only the approach by \cite{nr79} integrates a Vision Transformer (ViT). This model also leverages transformers for visual tasks, however compared to Swin Transformers, the input images are processed as sequences of patches without hierarchical divisions.
Document Image Transformer (DiT) \citep{dit}, which in turn is a self-supervised improvement over ViT, is also only used in the approach by \cite{nr58}. Transformer-based visual encoders might be less commonly used due to their higher computational and training complexity, despite their potential to capture richer visual features than CNN-based models.
The distribution of chosen decoders shows that the majority of approaches use sequence labeling layers for decoding. This usually consist of a linear layer and a Softmax layer, which assigns a probability distribution over all possible fields to each token in the sequence. The task of KIE is then performed by choosing the field with the highest probability according to the probability distribution.
Besides, some approaches alternatively use LSTMs \citep{lstm}, BiLSTMs \citep{bilstm}, Conditional Random Fields (CRF) \citep{crf} or a combination thereof. BiLSTMs are often used as they effectively capture contextual information in a bidirectional fashion. CRFs are useful, as they model the dependencies between labels in a sequence and incorporate a global context. These more complex models can therefore provide additional information compared to a simple Softmax layer and can potentially increase the extraction performance.
When looking at the decoder choice over time, one can see that the two aforementioned variants have not been used as frequently since 2022. Also, especially since 2023, newer decoding methods were examined. Nonetheless, sequence labeling layers remain popular over time. Some authors also use existing KIE approaches for decoding. For example, \cite{nr74} use the model by \cite{nr59} for their decoding step.

Overall, it can be said that there are many different approaches to encoder and decoder mechanisms, with no dominant approach. BERT-models and layout-aware variants like LayoutLM are most commonly used for textual encoding, while ResNet-based models are used as visual backbones. Decoding steps are usually performed either by simple sequence labeling layers or by more complex architectures that allow to incorporate certain dependencies within the token sequences.

\subsubsection{Graph-based methods}

The findings regarding the analyzed graph-based approaches are presented in table ST5 of the electronic supplementary material. A key property in which the approaches differ, is how the underlying graph representation is constructed. Most approaches use a fully connect graph, which means that every single node of the graph is connected to each other. As a result, the resulting graphs usually have a very large number of edges. 
Also common is the k-nearest neighbors algorithm, where each node has k neighboring nodes. The approaches use different values for k, ranging from 4 up to 100. Choosing the value for k is no trivial task and might introduce several biases \cite{nr99}. Also, increasing k will necessarily lead to more complex graph structures that include too many edges \cite{nr128}. As a result, KIE approaches may not be able to recognize the actual relationships within a document. Nonetheless, there is no study that investigates these aspects in depth.
Other common methods are to considering neighbors in the four major directions \textit{up, down, left} and \textit{right} or using the $\beta$-skeleton graph algorithm, which connects nodes based on their geometric proximity depending on the parameter $\beta$. To this end, all approaches choose $\beta$=1.
A few authors divide documents into distinct areas starting from each node (e.g., in 45 degree angles) and identify closest neighboring nodes in each of these segments. \cite{nr11} emit 36 rays out of each node and declare all other nodes, whose bounding box is crossed by these rays, as a neighbor.
In a few KIE systems, the graph creation is not defined by a heuristic algorithm, but rather iteratively learned by a neural network. The reasoning behind this choice is that the aforementioned heuristic algorithms are not always able to adequately represent VRD layouts.

Regarding different methods, how the information is propagated through the graphs, in most cases, a Graph Convolutional Network (GCN) is employed. GCNs propagate information between the graph nodes to iteratively learn representations, which capture both local and global contexts. Each iteration (i.e., graph convolution) calculates new embeddings for the nodes and edges. In some papers, a variation of a GCN defined on node-edge-node-triplets is used in conjunction with a Multilayer perceptron (MLP). In this setting, the features of the node itself, its edges and all neighboring nodes are fused in order to obtain the new node embedding.
Another common method for graph propagation is using Graph Attention Networks (GATs), either with our without multi-head attention. This architecture utilizes the attention mechanism based on transformers to calculate different weight coefficients for each neighboring node, allowing the model to focus on the most relevant information during graph propagation, and thus improving the overall node representations \citep{gat}. Yet, they are less used than GCNs, which might be due to the increased complexity and computational overhead of attention mechanisms, especially when dealing with large graphs.
With respect to the graph propagation, one can also analyze the number of propagation layers. The number of layers not only determines the overall system complexity, but also the receptive field of each node -- determining how far it can aggregate information from its neighboring nodes. Each layer increases the depth of this receptive field by one.
On average, the approaches consist of 3.7 layers. \cite{nr6} conducted ablation studies regarding different numbers of layers and found that the optimal number of GCN layers is 2, which however can be task-dependent and should be considered on a case-by-case basis.
With regard to the GAT-based approaches, one needs to decide on the number of attention heads. \cite{nr25} conducted a study with different attention heads and came to the conclusion that in their case, 26 attention heads produced the best results. The authors also mention that correlations between the number of attention heads and the number of fields to be extracted might exist. Here, too, it is necessary to examine which variant works best in each individual case.

The approaches differ in the way KIE is ultimately performed. Most of the approaches use a simple node classification, i.e., after the embeddings of each graph node have been obtained, they are used to classify each node into one of the fields to be extracted using a classification layer. Also common are BiLSTMs, CRFs and their combination in the form of BiLSTM-CRFs. Besides, multiple other methods are used to perform the KIE task. For example, \cite{nr60} use the Viterbi algorithm, which identifies the most likely sequence of states by iteratively calculating probabilities and backtracking through the graph. 

In by far the most cases, the graphs are defined on word-level, which means that every word on a document image represent one node in the constructed graph (see figure \ref{example_graph}). This is also inline with the granularities for the sequence-based methods. Another common way to define graphs is to consider entire segments or sentences. The resulting graphs are therefore rather coarsely defined. Some of the approaches consider entire text lines as nodes, regardless of whether the contained text elements are related to each other or not. A few authors define the graphs on more than one granularity simultaneously, both using fine-granular (e.g., words) and more coarse-granular (e.g., text segments) nodes.

Another central component of graph-based methods is the definition of node features. These features determine, which types of information can be utilized for KIE. In general, the node features usually consist of position-oriented values such as normalized coordinates of the node within the document image. Sometimes, relative distances to neighboring nodes are also part of the node features. 
Another common feature type is related to the textual content represented by the node, e.g., word embeddings. The approaches differ in terms of which embeddings models are used. In this regard, BERT-based-embeddings are often employed, which was also the observation regarding textual encoders of sequence-based approaches. Some other methods are Byte Pair Encoding \citep{bpe}, BiLSTM and LayoutLM. In some cases, multiple text embeddings are used simultaneously. For example, \cite{nr50} employ Word2Vec, BERT and LayoutLM for the word embeddings.
On top of positional and textual embeddings, many approaches also integrate visual features, often obtained from models such as ResNet or LayoutLM.
Many authors also consider hand-crafted miscellaneous features. The goal of these features is usually to incorporate certain node characteristics as well as relationships between nodes that should aid the model in performing KIE. \cite{nr7} for example include various boolean features whether a graph node represents a date, zipcode or known city, among others.
The approaches also differ regarding the size of the node feature vectors. Although in many cases the actual number is not explicitly reported in the manuscripts, on average, the features vectors have a length of around 500.  
In general, there is little discussion about embeddings sizes for node features in related work, from which it can be concluded that these have no significant influence on the performance of the KIE systems. Much more important seem to be the different types of features which are being integrated.
Another distinguishing factor of graph-based approaches is, whether they incorporate a global node in the defined graph representation. This global node can contain information about the entire document image and can be connected to the individual nodes of the graph. To this end, only the approaches by \cite{nr14,nr96} use such global nodes. \cite{nr85} also construct a global node, however it is not used as a information carrier and is rather formally required due to the chosen concept based on document layout trees.

Edges represent the second integral part of the graph structures. It is noticeable that a large portion of work does not incorporate any edge features at all. Consequently, less attention is paid to this type of feature compared to node embeddings. However, \cite{nr99} show that edge features, especially in terms of geometric data, can improve KIE performance, even more so than geometric information obtained from node features, as it integrates spatial information of the surroundings.
The edge-features are usually derived from pairwise positional relationships between the connected nodes. Most often, horizontal and/or vertical distances, both in absolute and relative values, are considered. In this regard, \cite{nr6} highlight the importance of the visual distances between two nodes. The aspect ratio of the bounding boxes that span the graph nodes is also often included as edge features. 
In only three of the graph-based approaches are visual embeddings part of the edge features.
Regarding the distribution of edge directions, i.e., whether the edges are directed or undirected, there is no clear dominant variant, however in the majority of the cases, the edges are undirected.
Additional research is required to appropriately estimate the impact of certain design features such as the edge direction on the extraction performance. To this end, \cite{nr103} show that both directed edges as well as introducing edge weights can improve extraction results. However, as illustrated before, these considerations are usually dependent on the underlying scenario and generalizing statements can only be made to a limited extent.

\subsubsection{Grid-based methods}
\label{grid}

As mentioned in section \ref{overview}, this paradigm is less popular in related literature and only 10 approaches make use of such grid structures. This suggest that while it offers an intuitive document representation, grid-based systems struggle with flexibility and generalizability across diverse and complex layouts of VRDs. The analysis of the approaches is shown in table ST6 of the electronic supplementary material.
An interesting observation is that most approaches represent evolutions of existing grid-based systems, in particular Chargrid \citep{nr2}. This shows that related research is mostly conducted within a very narrow corridor. Regarding the chosen granularities, a majority of the systems is defined on word level, while three are defined on character level and one method even on token level.

Table ST6 also lists findings regarding the grid dimensions. Almost all methods define the height and width of the grid based on the pixel counts of the input image. The grids are therefore rectangular, as this corresponds to the rectangular format of corresponding VRDs. \cite{nr4} and \cite{nr66} use a fixed value for the height and width, namely 512 and 336 respectively, therefore resulting in square grids. \cite{nr33} scale down the input dimensions by a factor of 8 in order to reduce the overall model complexity. The work of \cite{nr48} uses a somewhat different approach and employs a dynamic grid structure, in which the height and width dimensions are defined by the distance between the maximum and minimal coordinates of bounding boxes in vertical and horizontal directions -- i.e., the span of actual document content.
Another distinguishing factor is the feature vector length. Relatively similar values are chosen here, with 256 and 768 being the most common vector sizes. The approach of \cite{nr4} uses the smallest feature set with a vector of length 35. The most complex feature vector is defined by \cite{nr8}. In this case, the feature vector has the dimension 4×128×103.

Regarding which features are chosen for the individual grid elements, the majority of approaches use text embeddings obtained from different models -- mostly BERT-based architectures. Another common feature type is a 1-hot encoding of the characters, which means that each character in the word corpus is assigned to a specific index value.
A few approaches also integrate visual features based on pixel-wise RGB-values, ResNet18, or Swin Transformer embeddings.
Another distinguishing feature is the integration of features for background elements of the grid, i.e., all elements that are not overlapped by contents of the document image. In this regard, almost all approaches use an all-zero feature vector. The aim is to clearly differentiate actual content and backgrounds. \cite{nr8} use a sparse tensor and therefore discard background elements entirely. Only the work of \cite{nr33} integrate features for the background elements, namely the RGB-channels of the corresponding coordinates in order to fuse visual information into the architecture.

Most of the methods employ semantic segmentation to perform the KIE task, where each element of the document image is assigned to semantic categories (i.e., fields to be extracted), which results in a segmentation mask of the original document image.
Other choices are bounding box regression, where the goal is to predict bounding boxes of semantic entities, which can be helpful to better differentiate individual objects. For example, when processing an invoice, it is helpful to separate each individual line item. This is also the reason why a few approaches combine semantic segmentation with methods for bounding box regression (or line item detection in general) \cite{nr2,nr3,nr66}.

\subsubsection{LLM-based methods}
\label{llm}

With the emergence of LLMs across many research areas, their application has also been studied for KIE in the recent past. Multimodal LLMs in particular have accelerated this process, as they can adequately process VRDs based on different modalities. Due to the nature of the models, KIE is typically formulated as a VQA task. That is, the LLMs receive OCR text as input and are prompted with questions like \textit{''What is the value for the \{key\}?''}. All LLM-based approaches are generative by design. A key advantage of these approaches lies in the basis of LLMs: they have been pre-trained on extremely large amounts of data, significantly larger than those used in KIE-specialized sequence-based models (see also section \ref{sequence}). This allows them to draw on extensive world knowledge, which can result in a higher generalization capability compared to more traditional KIE methods. The analysis of LLM-based approaches is provided in table ST7 of the electronic supplementary material.

More than half of the approaches use an external OCR engine. This shows that robust text recognition remains important and should not be left solely to the LLM itself. This point is underlined by the fact that only \cite{nr92,nr101,nr113}  operate purely on the image modality.
Almost all methods use open-source LLMs as a backbone, with LLama2 being particularly common. In most cases, a visual encoder is used to integrate the visual information. To this end, three of the approaches use LayoutLMv3 for image encoding. The works vary in terms of image resolution. Some use high resolutions such as 2560×2560, while others use lower resolutions like 224×224 but compensated by extracting up to 20 image crops in order to more precisely capture individual document regions. Here, a trade-off must be made between computational overhead and extraction quality.

Some approaches perform continued pre-training to better adapt the backbone LLMs to VRDs. Tasks used during this phase include text recognition, text spotting, and table understanding. This continued pre-training is often carried out on existing KIE datasets, such as IIT-CDIP. With the exception of \cite{nr80}, all approaches additionally perform supervised fine-tuning (SFT) for KIE and therefore adjust model weights. The goal is to improve the foundation models' capabilities of processing VRDs and perform KIE. For this purpose, corresponding instruction-tuning datasets are created, often based on popular benchmark datasets such as CORD. This involves converting the document annotations into question-answer pairs similar to the VQA-style question mentioned before. Besides, some authors also include various other SFT tasks such as image captioning where the LLM is prompted to generate textual captions for input documents. 
In this regard, very few approaches use parameter-efficient fine-tuning methods such as LoRA. Instead, most train all parameters of the LLMs during SFT. This indicates that in order to properly fine-tune the foundation models to VRDs and KIE, it is required to update the weights of the majority of the architecture. However, there is no work that analyzes this aspect in depth.

The approaches also differ in how the different modalities are fed to the LLMs. For example, \cite{nr80} pass OCR data to the LLM and prompt it to obtain the corresponding class. An example format is: \textit{''Q3:\{text: 'TO:', Box:[102 345 129 359]\}..., What are the labels for these texts?''}. Incorporating layout information explicitly by adding a list representation of the bounding boxes into the prompt like above is a straightforward and commonly used approach. However, it is associated with limitations. Using this method, a large part of the input sequence is therefore occupied by these tokens, which reduces the available context length for the actual VRD content. Also, the backbone LLMs are typically not (pre-)trained with such prompt formats. This can lead to the model struggling to properly reason about the layout information.

\cite{nr116} take a different approach by converting documents into an HTML structure and passing it to the LLM in order to obtain a structured JSON output with the extracted information. The authors argue that LLMs are well-suited to processing HTML, as a large portion of their pre-training data consists of HTML. 
Besides, the common procedure of LLM-based approaches is to obtain image encodings and pass them to the LLM together with the textual prompt to get the extracted value for the respective field. 
Some works include all fields to be extracted in one prompt (e.g., \cite{nr116}), while others only consider one field per call (e.g., \cite{nr92}). The latter requires multiple calls per document, which can increase runtime, but may allow the LLM to better focus on one field and improve extraction quality per field.
Authors also experiment with different prompt templates. For example, KIE is sometimes framed as a multiple-choice task: \textit{''\{document\} What is ‘\{value\}’ in the document? Possible choices: \{keys\}''}, where \textit{\{keys\}} is a randomly selected subset of key names from the dataset \cite{nr130}. The idea is to leverage the LLM’s prior knowledge of field names to better identify the correct value.
In general, most works follow similar prompt formulations. However, \cite{llm_templates} shows that it is advantageous to create instruction datasets with heterogeneous prompts, as this increases robustness for different layouts of VRDs.

\subsubsection{Other methods}

Of the 130 papers analyzed, seven cannot be assigned to any of the paradigms of KIE methods discussed before. In the following, we briefly present key concepts behind those deviating approaches.
The approach by \cite{nr16} is based on learning representations of document snippets and classifying them into fields to be extracted. First, potential candidates are identified for each field based on the data type. 
The next step is to select the correct candidate for each field. To this end, a representation is obtained for each candidate as well as each field to be extracted. The candidate representation is constructed based on the candidate itself as well as its neighboring segments, utilizing textual and positional information processed by a self-attention mechanism. 
Finally, a similarity score is calculated for each pair of candidate embedding and field embedding. The similarity scores are then used in a separate module to select the appropriate candidate for each entity, which can be defined in a variety of ways. A trivial method is to select the candidate with the highest similarity score for each entity.

\cite{nr19} use a hierarchical tree-like structure to represent fragments of the document pages, similar to the graph-based methods. Compared to traditional tree architectures, nodes on the same hierarchy level can also be connected with each other.
Parent and child nodes of these trees represent key-value pairs. 
The KIE task is performed by predicting the relations between the individual fragments, i.e., directions of the edges. Obtaining the most likely parent element of each fragment can be ultimately used for identifying fields to be extracted.

\cite{nr23} propose an end-to-end KIE system and combine text reading and text understanding.
The text reading steps includes a detector model as well as an LSTM-based recognition model which identify text in the document images. The approach also includes a dedicated module to fuse the different input modalities, which is then used by the final KIE module. A BiLSTM coupled with a fully connected network is employed as a decoding mechanism to predict relevant entities for the fields of interest. This approach is therefore closely related to sequence-based KIE methods.

\cite{nr44} follow concepts of \cite{nr16} in the sense that for each field to be extracted, a set of candidate spans is identified first. In order to identify candidate spans for a named entity, multiple detector functions are implemented, to some extent based on domain knowledge. The authors then employ an adversarial neural network in order to find local contexts of the identified visual spans. These local contexts represent specific fragments of a document image that contain the spans as well as related relevant context (e.g., a larger paragraph). Based on these identified segments, both global and local context vectors are constructed with textual and visual features. These features are then used by a binary classification model in order to predict, whether a visual span contains a specific field to be extracted or not.

The approach by \cite{nr46} is another method that is based on generating candidates for each field and scoring them subsequently. Given a document image and fields to be extracted including their data types, the system first identifies candidates in the OCR-output based on third-party detector functions. The identified candidates are then scored according to their likelihood of correctly representing a particular field of interest in a binary classification setting. Similar to \cite{nr16}, the score depends on the similarity between the embeddings of the candidate and the field of interest. 
Different scoring functions can be integrated, however the authors choose a function that assigns the candidate with the highest similarity score to each field.

\cite{nr88} propose a novel document representation structure which they refer to as cell-based. This representation is similar to grid-based methods, however the cell-based methodology has no consistent placement of elements in terms of height and width. Instead, the cells are defined depending on the actual document content. For example, different lines or columns in the cell-structure can consist of a different number of elements. The individual cells are also sorted by row and column index respectively, which provides additional information to the KIE system. The obtained cell-based layout is then processed by sequence-based methods such as LayoutLM and therefore follow the typical sequence labeling scenario that respective models use to perform KIE.

Instead of processing entities sequentially or constructing elaborate graph structures, the approach by \cite{nr112} constructs a token pair representations matrix based on multi-modal features from a pre-trained encoder. It then jointly generates three relation matrices: line extraction (to determine the boundaries of text lines), line grouping (to merge related lines that form multi-line entities), and entity linking (to connect keys with their corresponding values). These predictions are then combined to ultimately obtain extracted key-value pairs.

\subsection{Evaluation}
\label{eval}

\subsubsection{Methodology and setup}
\label{eval_methodology}
Table ST8 of the electronic supplementary material shows the findings regarding the presented evaluation setups. A consistent picture throughout is that authors often do not describe their evaluation procedures in great detail and it is also not always obvious from the manuscript itself. This is the case for around half of the analyzed papers. We declared affected properties as ''n/a''.
In 19 of the analyzed papers, an element-based evaluation method is chosen, which means that elements like the predicted class of a token are compared with the groundtruth class in case of sequence-based approaches. Another common element-based evaluation is to compare the predicted bounding box with the actual bounding box and to identify a match or mismatch based on the overlap of their respective coordinates. 
45 of the analyzed approaches incorporate a string-based evaluation setting. In such cases, the extracted textual values are compared with the groundtruth strings to determine, whether a prediction for a given field was correct or not. A string-based evaluation is particularly relevant for assessing the suitability for real-world applications where the extracted texts are used for further processing (e.g., transfer to other information systems).

In almost all of the manuscripts, the authors present the overall performance, i.e., the aggregated results across all fields of interest and all documents of the test set. However, the field-level performance, which shows the performance for each field individually, is only adopted in around 30\% of the analyzed papers. 
One aspect that is almost never presented, is the performance on the \textit{Unknown} class. In the context of KIE, \textit{Unknowns} represent all elements of a document that do not belong to any of the fields to be extracted. This can be helpful in determining the extent to which an approach is able to distinguish irrelevant content from relevant parts of a document image.
In most cases, the proposed method is compared against existing KIE systems, either by comparing the own results with the evaluation metrics presented in the respective papers or by re-implementation. Thus, it can be stated that a high emphasis is placed on the overall comparability and the legitimization of the own approach. However, especially when authors decide to re-implement existing approaches, comparisons of evaluation results are not always appropriate, since less effort is usually put into training the approaches of third parties.

Considering the employed evaluation metrics, one can see that the F1 score is used most of the time. Precision and Recall, with the F1 score being the harmonic mean of these two metrics, are only explicitly reported in about 25\% of the cases. Interestingly, the usually widely used Accuracy metric is only presented by a few authors. One reason for this is that KIE datasets are usually very unbalanced in terms of label distribution, which strongly distorts the overall Accuracy metric \citep{accuracy} -- making it less representative for these scenarios.
Nonetheless, given that KIE is usually seen as a multi-class classification problem, it is not surprising that authors adopt metrics such as Precision, Recall and F1, which are very common for such tasks.
As mentioned before, many authors present a string-based evaluation. To this end, various different metrics are used, e.g., Word Accuracy Rate, which counts the number of substitutions, insertions and deletions between groundtruth and predictions \cite{nr33}.
Moreover, it is noticeable that a large majority of the identified evaluation metrics are custom-defined and only used in one particular paper, which prevents an adequate benchmark against existing KIE approaches.

Custom datasets are being used in 54 of the analyzed papers and typically represent in-house document collections that are not publicly available. However, in around 80\% of the cases when custom datasets are used, the authors additionally present their results on public benchmark datasets to allow for a comparison with other KIE systems. In addition, there are numerous datasets that were proposed in individual papers, but were not adopted in any other subsequent work. This indicates again that KIE literature strongly focuses on a small set of benchmark datasets.

\begin{table}[!ht]
\caption{Common datasets for KIE research}
\label{kie_datasets}
\begin{minipage}{\textwidth}
    \begin{tabular}{llccccc}
        \toprule
        \multicolumn{1}{c}{\textbf{Dataset}} & \multicolumn{1}{c}{\textbf{Documents}} & \textbf{\# docs} & \textbf{\# classes}\footnote{in terms of entities to extract} & \multicolumn{1}{c}{\textbf{Language}} & \textbf{\# uses}\footnote{as part of analyzed approaches} & \textbf{Pre-Training?} \\
        \midrule
        FUNSD & Forms & 199 & 4 & eng & 61 & \ding{55} \\
        CORD & Receipts & 1,000 & 30 & eng & 52 & \ding{55} \\
        SROIE & Receipts & 973 & 4 & eng & 45 & \ding{55} \\
        IIT-CDIP & Lawsuits & 6,000,000 & / & eng & 24 & \ding{51} \\
        XFUND & Forms & 1,393 & 4 & zho,jpn,spa,fra,ita,deu,por & 15 & \ding{55} \\
        EPHOIE & Exams & 1,494 & 10 & zho & 9 & \ding{55} \\
        RVL-CDIP & Lawsuits & 400,000 & / & eng & 7 & \ding{51} \\
        Kleister-NDA & NDAs & 540 & 4 & eng & 6 & \ding{55} \\
        Kleister-Charity & Reports & 2,778 & 8 & eng & 6 & \ding{55} \\
        \bottomrule
    \end{tabular}    
\end{minipage}
\end{table}

Table \ref{kie_datasets} provides an overview of the most common datasets adopted in KIE research. Listed are benchmark datasets as well as pre-training datasets.
By far the most common datasets in this regard are FUNSD \citep{FUNSD}, CORD \citep{CORD} and SROIE \citep{SROIE}, all of which are used by at least a third of the papers. FUNSD includes forms from various domains. CORD and SROIE on the other hand contain photographed receipts, mostly from supermarkets and restaurants. 
There are discrepancies both in terms of dataset size and in the number of keys to be extracted. The datasets with many hundreds of thousands of documents are typically used for the pre-training of corresponding models and not as an evaluation benchmark. Therefore, these datasets are also widely adopted in other DU tasks.
Two of the three most used datasets only aim at extracting four fields of interest, which is a relatively low amount compared to the variety of information corresponding documents usually include. CORD on the other hand includes 30 fields to be extracted, which is also the highest among the listed datasets. The key difference between CORD and SROIE is that in case of former, detailed information including individual line item attributes such as their quantity or unit price need to be extracted, while SROIE only requires to extract aggregated information such as the total price.
For the most part, the datasets include English documents. One exception is XFUND \citep{q20}, which specifically intends to investigate multi-lingual capabilities of KIE approaches. Some popular Chinese datasets exist, namely EPHOIE and Ticket, however they are understandably not as widely adopted as most English counterparts that are being used by the international research community.

\subsubsection{Quantitative comparison}
\label{quantitative_comparison}

To ensure a meaningful and representative quantitative comparison, in the following we focus on the by far most commonly used datasets CORD, FUNSD and SROIE. We also only consider the F1 score, as it is the most commonly used evaluation metric. These choices allow for the largest possible sample size and a comparability across a wider range of approaches. Note that we do not distinguish between micro, macro or weighted F1 averages\footnote{For an explanation of these aggregated metrics, we refer to \cite{metrics}.}, since in many cases it has not been explicitly specified by the authors. The results should therefore be treated with caution.

\begin{figure}[!ht]
    \centering
    \begin{subfigure}[b]{\textwidth}
        \centering
        \begin{minipage}{.49\textwidth}
            \centering
            \includegraphics[width=\linewidth]{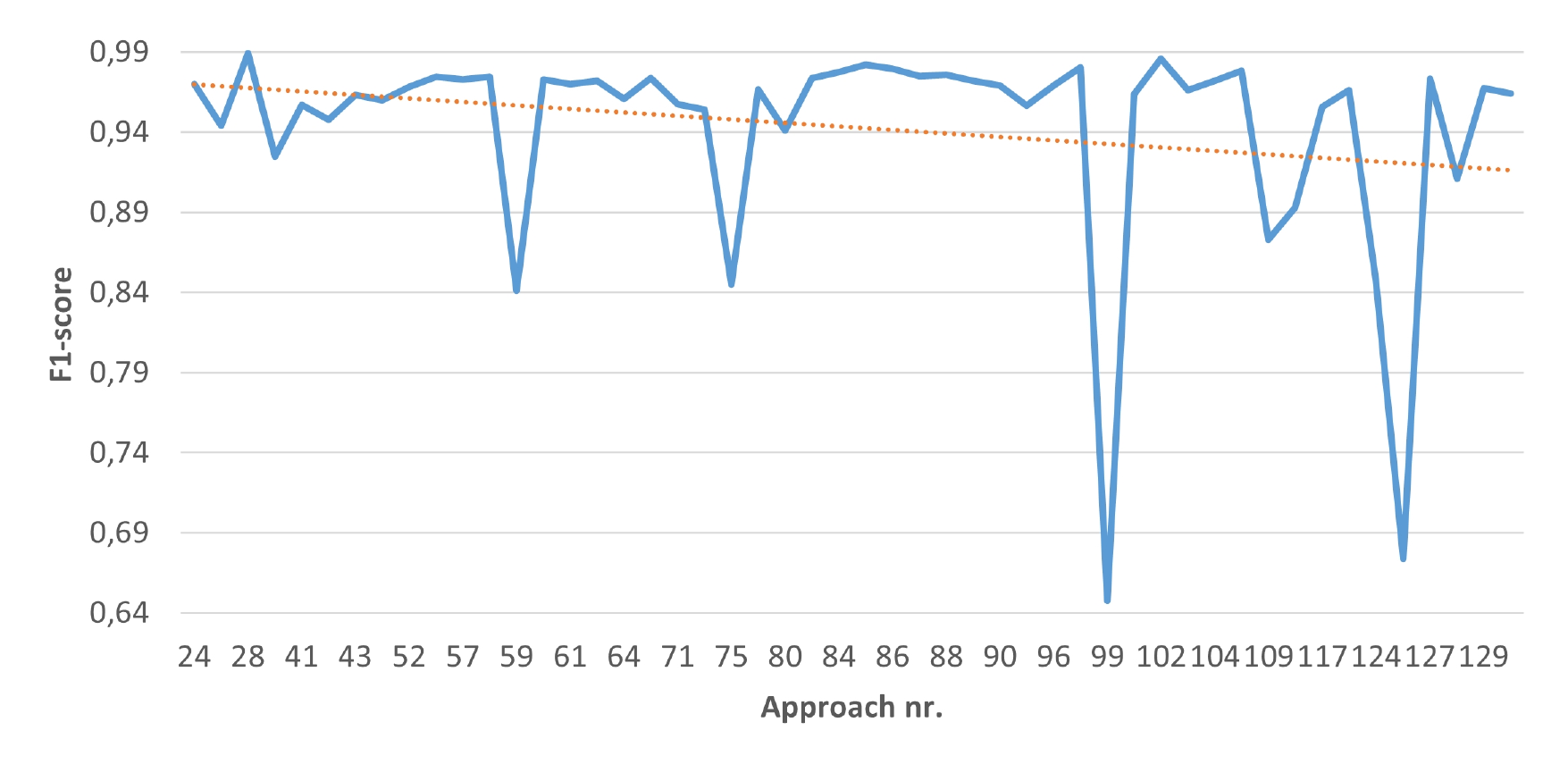}
        \end{minipage}%
        \hfill
        \begin{minipage}{.49\textwidth}
            \centering
            \includegraphics[width=\linewidth]{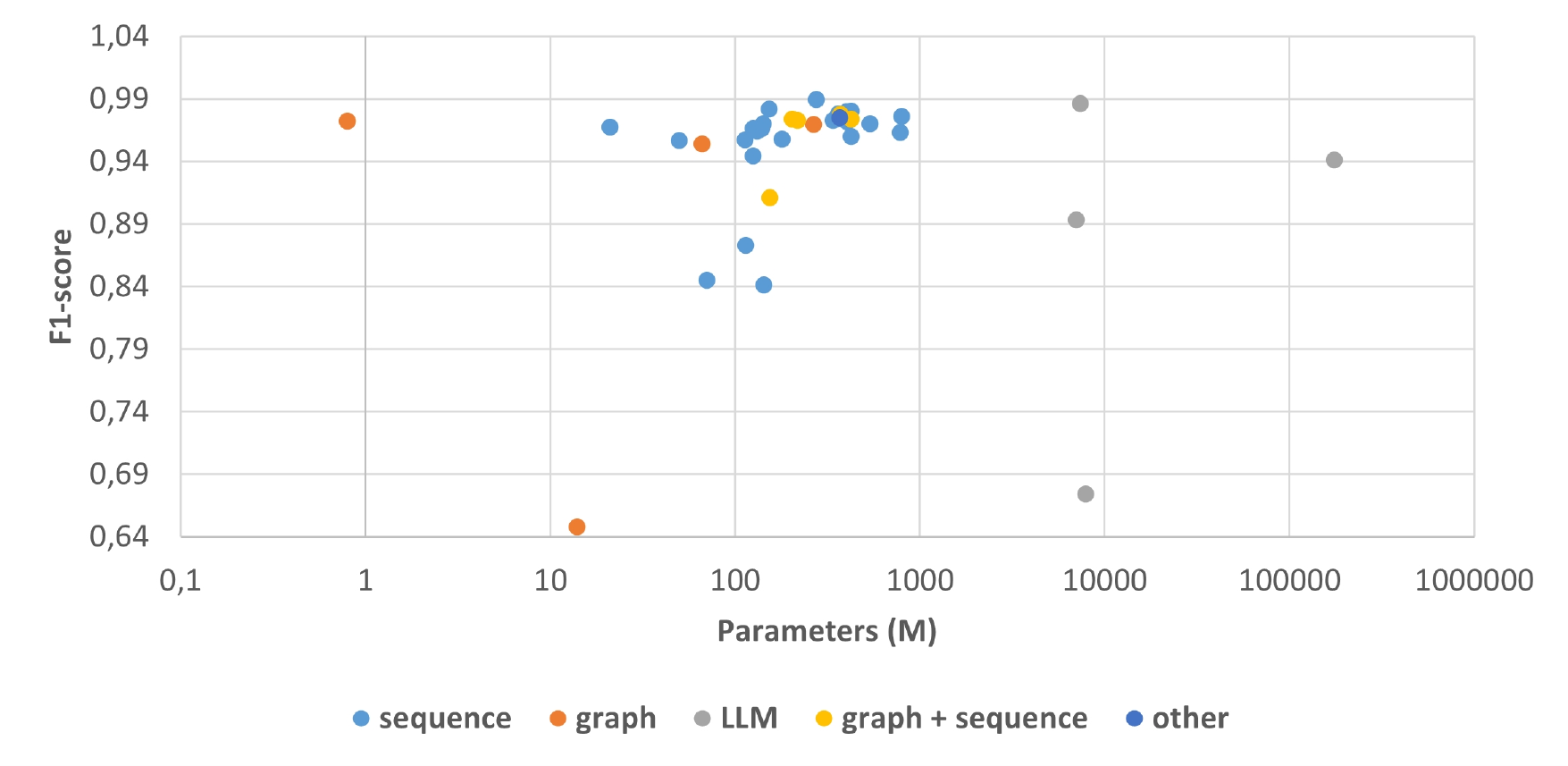}
        \end{minipage}
        \caption{CORD}
    \end{subfigure}
    
    \vspace{1em}
    
    \begin{subfigure}[b]{\textwidth}
        \centering
        \begin{minipage}{.49\textwidth}
            \centering
            \includegraphics[width=\linewidth]{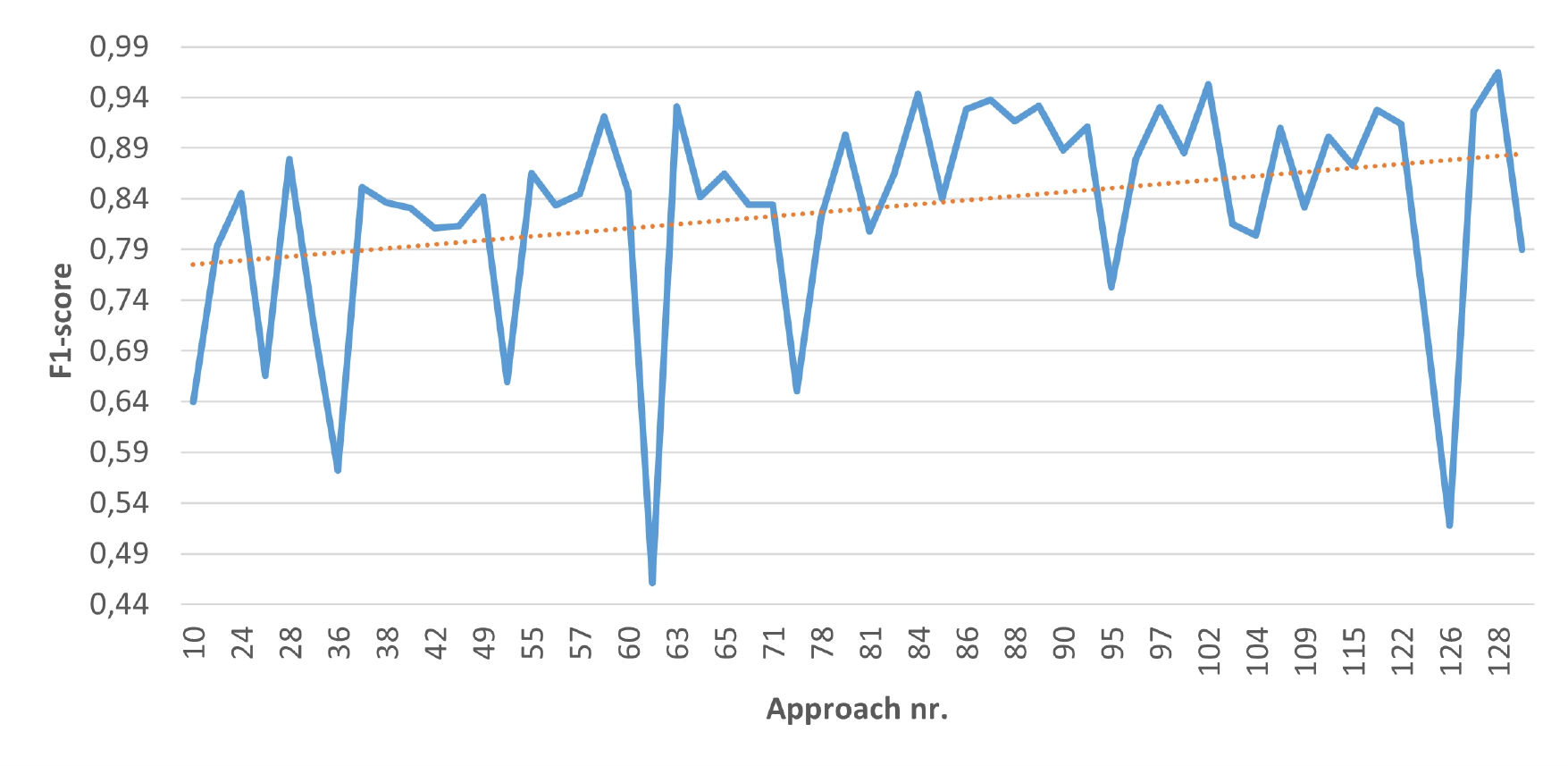}
        \end{minipage}%
        \hfill
        \begin{minipage}{.49\textwidth}
            \centering
            \includegraphics[width=\linewidth]{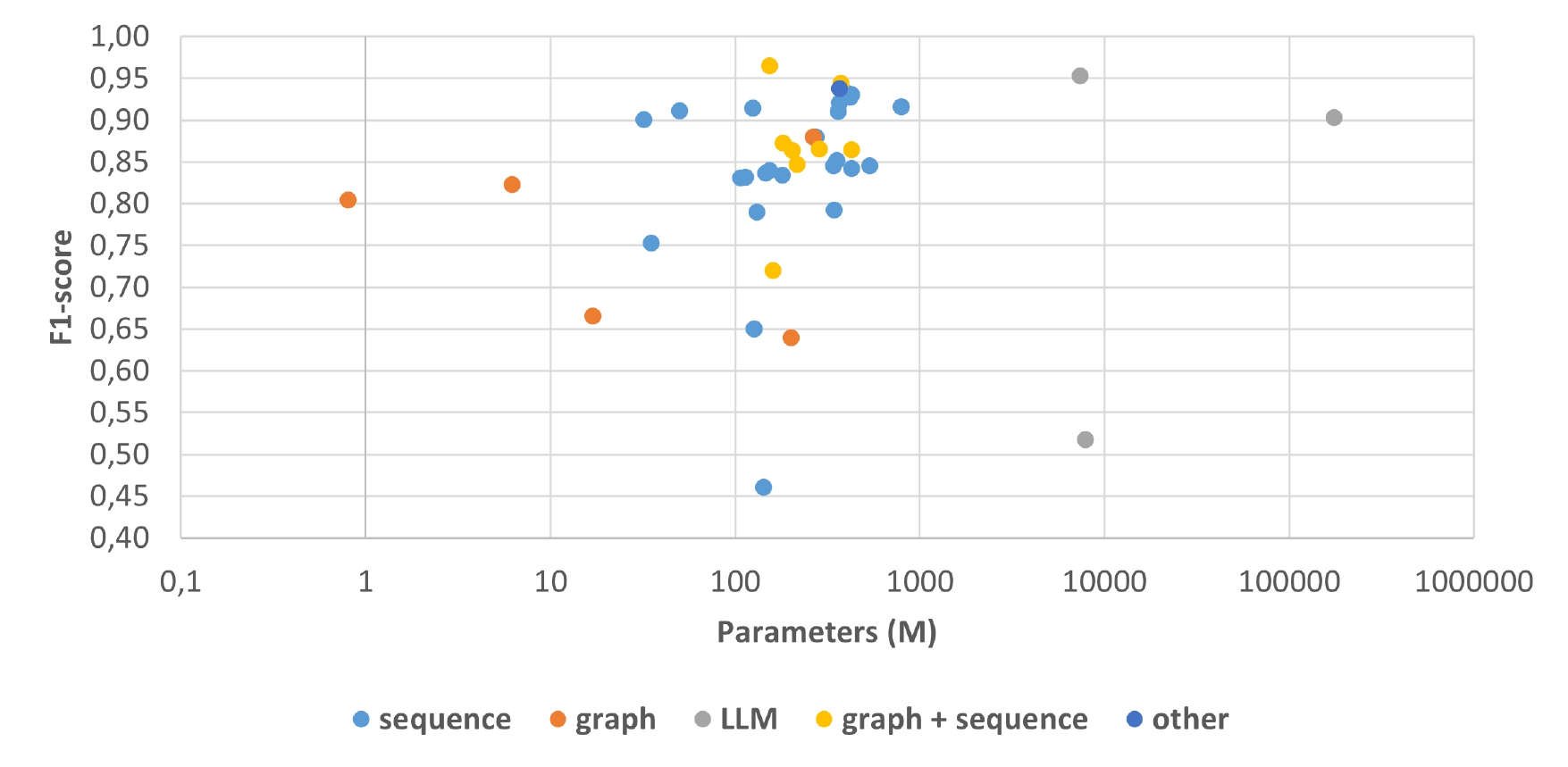}
        \end{minipage}
        \caption{FUNSD}
    \end{subfigure}
    
    \vspace{1em}
    
    \begin{subfigure}[b]{\textwidth}
        \centering
        \begin{minipage}{.49\textwidth}
            \centering
            \includegraphics[width=\linewidth]{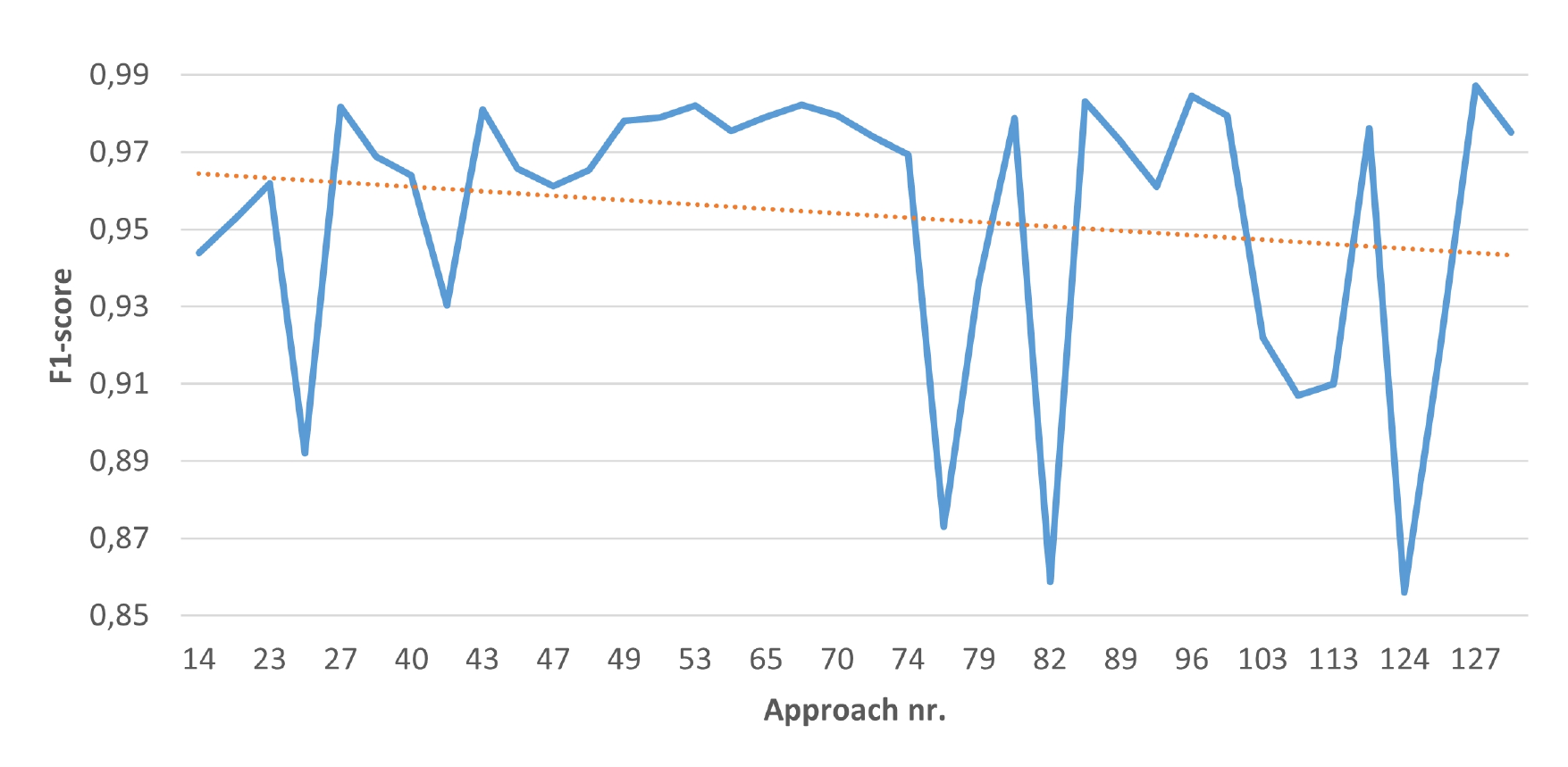}
        \end{minipage}%
        \hfill
        \begin{minipage}{.49\textwidth}
            \centering
            \includegraphics[width=\linewidth]{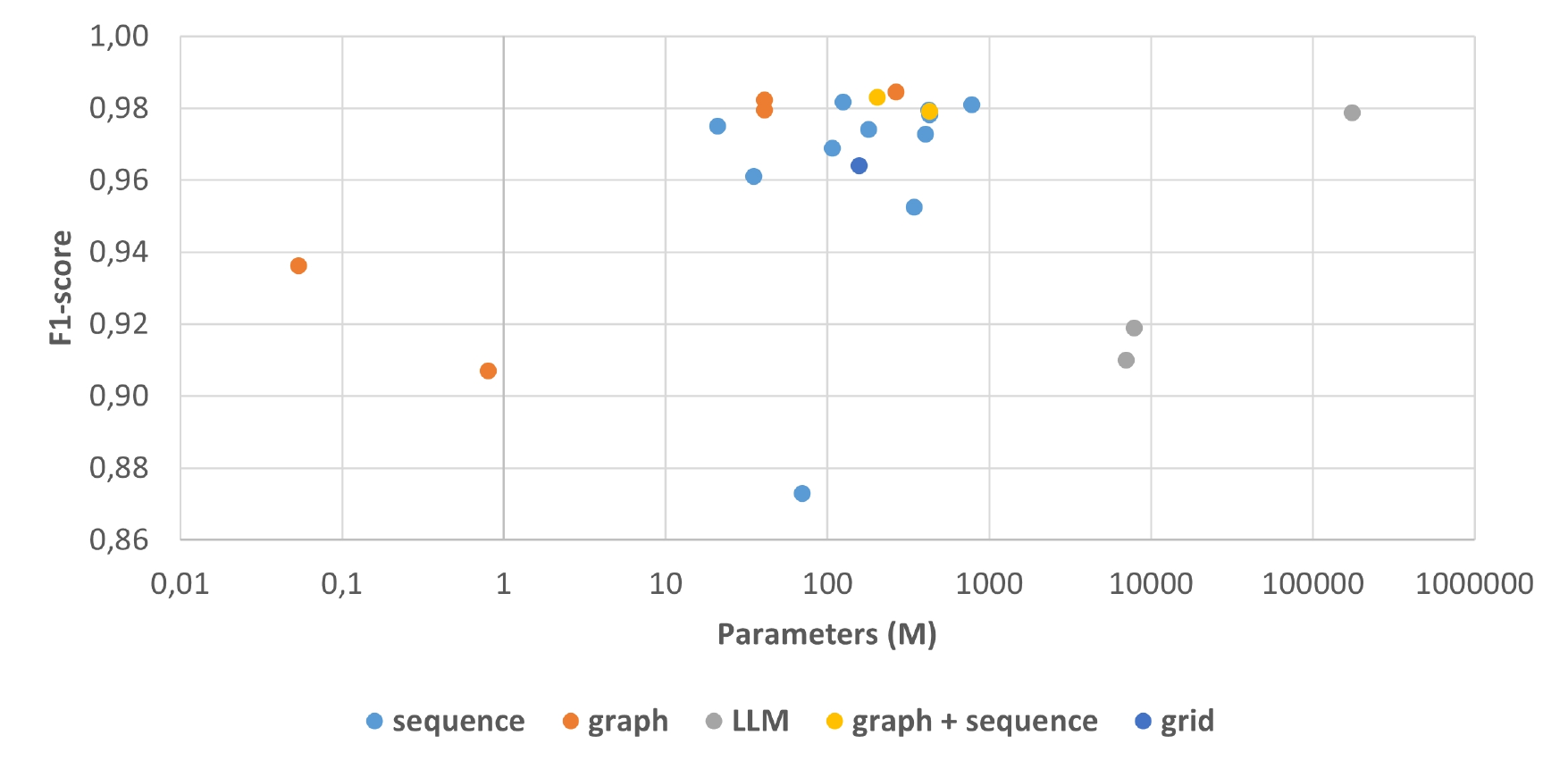}
        \end{minipage}
        \caption{SROIE}
    \end{subfigure}
    \caption{Evaluation results and correlation between model size and performance}
    \Description{Graphs that show the performance on three benchmark datasets and the fact that there not necessarily exists a correlation between model size and performance.}
    \label{correlation}
\end{figure}

To this end, figure \ref{correlation} shows the results for the three datasets given all papers that present results on the corresponding datasets. A trend line is also displayed. Only FUNSD shows an improvement over time, while for the other two datasets the trend line shows a decline. However, all graphs show a few outliers that performed worse than previous and subsequent work. One reason for this could be a different calculation of the F1 scores, e.g., by using macro averages instead of micro averages. Another possible factor could be the alignment of the models with the specific nature of the task. For example, models that do not take into account the spatial and semantic features of the datasets may perform poorly despite having a large number of parameters. These outliers will therefore negatively affect the trend line. The best performing approach for CORD is proposed by \cite{nr28}, for FUNSD by \cite{nr128}, and for SROIE by \cite{nr127}.
On average, very comparable results are obtained for CORD and SROIE, with an average F1 score of 0.94 and 0.95 respectively. This can be attributed to the fact that both datasets consist mainly of structured receipts, where layout and spatial encoding play a crucial role in performance. In contrast, FUNSD shows significantly worse results, with an average F1 score of only 0.83. One problem that KIE approaches face with FUNSD may be that this dataset was not primarily constructed for the KIE task. The fields to be extracted often span multiple lines of text, which is very different from traditional KIE datasets that aim at extracting key information such as a specific date.

Figure \ref{correlation} also shows the relationship between parameter count and performance across the three datasets, with the x-axis in logarithmic scale. In all cases, there is a clear clustering with some outliers. The correlation coefficients are -0.02 (p=0.90) for CORD, 0.08 (p=0.60) for FUNSD, and 0.11 (p=0.63) for SROIE, indicating no significant correlation between model size and performance. Notably, the slight negative correlation for CORD suggests that smaller models can outperform larger ones.
The figure also distinguishes KIE paradigms by color. Notably, graph-based methods often appear as outliers. Especially in case of FUNSD, this observation can be attributed to its unique layout, where entities often span multiple lines, making graph representations suboptimal. Hybrid models combining graph and sequence-based approaches tend to outperform pure graph-based models, especially in this scenario.
The analysis further shows that increasing model size beyond a certain threshold yields diminishing returns. This suggests that architecture design, tailored to the task, is more critical than model size. This insight is particularly useful in resource-constrained environments, where smaller, task-aligned models can achieve competitive or superior results.
The performance spread in FUNSD is wider compared to CORD and SROIE, with some smaller models struggling, likely due to the complex layout and semantic requirements of the dataset.
Interestingly, the LLM-based approaches often do not achieve competitive results and in some cases are among the worst performing KIE systems. Only the LLM-based approach by \cite{nr102} achieves almost state-of-the-art results for CORD and FUNSD. This highlights again that larger models don’t always lead to better performance if the architecture is not suitable. On the contrary, the approach by \cite{nr104} with only 0.8 million parameters achieves promising results and is even one of the best systems in case of CORD.

In conclusion, this analysis emphasizes that model design and task alignment are more important than model size. For structured documents as seen in CORD and SROIE, simpler models with strong layout awareness are effective, while hybrid models combining sequence and graph-based features work best for more complex KIE settings like FUNSD. Architectural choices need to balance semantic complexity with layout structure to optimize KIE performance.
It is important to note that the findings are based on datasets with strong domain overlap -- in particular CORD and SROIE, which consist of receipts -- and that all three analyzed datasets are in English. As a result, the findings may have limited generalizability to other application domains or languages. The outlined performance trade-offs may be more pronounced for other document types and layouts, where larger models could yield significantly better results.

\section{Discussion}
\label{discussion}

\subsection{Technical analysis}
\label{qualitative_comparison}

KIE approaches vary widely in how they represent and process VRDs, with strengths and weaknesses in handling multi-modal inputs, spatial document structures, and semantic richness. 
Sequence-based models, such as LayoutLM and its variants, dominate due to their ability to leverage large-scale pre-training and various attention mechanisms. These models incorporate spatial information via positional embeddings and often use linear classification layers to perform NER. However, such classification layers are limited in modeling label dependencies across sequences -- unlike CRFs -- and struggle with discontinuous entities, which are common in real-world documents \citep{nr123}. Moreover, while end-to-end models provide a unified extraction pipeline, OCR-dependent systems still achieve superior performance in benchmarks since they decouple the complex tasks of text recognition and entity extraction \citep{e2e_vs_ocr}.
Graph-based models excel at capturing document structure through spatially and semantically defined graph structures. At the same time, GCNs and GATs allow for effective layout-aware propagation. However, many approaches rely on heuristic graph construction, which can introduce noise or bias. In contrast, learned graph structures can show improved robustness and generalization at the expense of model complexity \citep{nr128}. Edge features remain underutilized, despite evidence of their potential to improve spatial reasoning.
Grid-based approaches treat documents as grids and apply segmentation-like techniques for KIE. These methods are particularly suited to highly structured VRDs, but are less flexible in adapting to different layouts. They also tend to use limited semantic encoding, with little advancements beyond early design choices such as those proposed by \citep{nr2}. This is also reflected in the fact that research on grid-based approaches was mostly conducted before 2022.
Novel LLM-based methods introduce a generative and instruction-driven paradigm for KIE. While these models offer strong generalization and task flexibility through their use of large foundation models, they have practical limitations. Layout information is often injected via simple prompt engineering, which is not native to the LLM training procedure and can impact efficiency. Furthermore, inference times are significantly higher - under certain circumstances up to 120 times higher than other approaches on comparable hardware \citep{nr117} — which poses scalability challenges. Although hallucinations (i.e., the generation of text not present in the original document) are rare, occurring less than 1\% in some cases \citep{nr117}, dedicated mechanisms may still be required to prevent them.

Almost all KIE approaches have limited capabilities to handle long and multi-page VRDs. This is mainly due to architectural constraints, e.g., transformers have a fixed input sequence length, usually 512 tokens, which is often not even enough to capture even a single page of a VRD. Graph-based and grid-based methods are unable to capture multi-page dependencies because their document representations are defined page by page. LLM-based approaches can overcome this problem to some extent, as the backbone LLMs usually provide longer context lengths (e.g., 4,096 tokens). This aspect is less of a problem when processing VRDs with fewer words (e.g., receipts as seen in CORD and SROIE), but can play a role in documents with more content, such as forms (see FUNSD). All in all, an adequate understanding of long VRDs and maintaining context over long sequences remains a challenge. 

Overall, the different categories require trade-offs between semantic expressiveness, spatial reasoning, and practical constraints. Transformers offer strong contextual understanding but need improvements in sequence modeling and layout perception. Graph-based methods capture document structure explicitly and can benefit from learning the graph connectivity, rather than relying on heuristic algorithms. Grid-based models handle layout well, but lack semantic flexibility. LLM-based approaches allow for a broad adaptability, but remain computationally intensive and less effective at precise spatial reasoning.

\subsection{Pre-Training and Fine-Tuning procedures}
\label{train_setups}

An increasingly popular approach in KIE research, in particular when dealing with sequence-based methods, is to split model training into self-supervised pre-training and supervised fine-tuning steps. The aim of the former is to acquire a general document understanding by the model first, which can then be utilized for different downstream tasks such as KIE as part of subsequent fine-tuning steps.
In general, pre-training procedures require large volumes of data to properly let models learn corresponding document representations, which is why most authors use large datasets such as IIT-CDIP, as also discussed in chapter \ref{sequence}. The problem with this dataset in particular is that included documents originate from the 1990s and therefore show poor image quality, noise produced by faulty scans as well as a generally low image resolution.
This also means that a large proportion of the proposed approaches are pre-trained on outdated documents and may therefore have difficulties when they are fine-tuned on newer documents as part of real-world use cases. Another aspect that should be considered is the fact that all documents of IIT-CDIP are related to lawsuits against the tobacco industry and therefore lead to a strong domain bias that is being adopted by the models during pre-training. In this respect, it is debatable whether it would not be more effective if pre-training procedures were carried out based on more representative datasets (see also chapter \ref{agenda}).

As discussed in chapter \ref{sequence}, the majority of authors adopt MVLM during pre-training. In some cases, it is even the only pre-training objective that is being used. Given the general popularity of MLM in the NLP domain, it is not surprising that the MVLM task, which is derived from it, is also popular in KIE research. It represents an effective method to let models learn useful representations given the surrounding textual and positional contexts.
As the adequate processing of VRDs also requires to consider visual cues, newer pre-training objectives have been proposed in order to fuse visual information into KIE systems. This can be done in different ways, for example by letting the model reconstruct local or global document image snippets \citep{nr24, nr89} or predicting the segment length of an image snippet \citep{nr39}. Other possibilities aim at various matching-tasks between text and document images, for example where the model has to predict whether a sentence describes a document image \citep{nr24}, if a given image and text are part of the same page \citep{nr49} or if image patches of a word are masked or not \citep{nr58}.
Some authors also define additional pre-training objectives to better exploit positional information, for example by the means of classification problems to estimate a tokens placement into a document image area \citep{nr37, nr64} or to estimate relative positional directions of an element to its neighbors \citep{nr39}.
Research is also conducted related to a better understanding of numerical values and their relationships in business documents \citep{numeric_ordering}. This can be beneficial for processing document types such as invoices, which usually contain various monetary values that are closely related to each other.
To conclude, the proposed pre-training objectives are carefully designed to allow for an adequate fusion of textual, positional and visual inputs, which helps corresponding models to obtain a document understanding which can then be exploited in downstream tasks like KIE.

Another category of pre-training objectives stems from the generative KIE methods, where corresponding models need to learn both natural language understanding and natural language generation (i.e., generating output sequences conditioned on input sequences) capabilities. To this end, \cite{nr52, nr53} adopt the pre-training tasks defined by \cite{UniLM}, namely Unidirectional LM, Bidirectional LM and Sequence-to-Sequence LM. Importantly, all pre-training tasks are considered simultaneously by using a shared transformer network which can alternate between the three objectives.
\cite{nr89} propose a range of self-supervised and supervised generative pre-training objectives based on task prompts and target outputs in textual form, for example where the model must predicts missing texts and locate them in document images as a structured target sequence.
The approach by \cite{nr72} follows a specific strategy for learning text recognition, document understanding and generative capabilities. It generally stands out in terms of its pre-training setup, as over 25 different objectives across different document types are being used. Nonetheless, a strong focus is placed on MLM-related tasks. One important pre-training task revolves around the model predicting a structured JSON output for a given document, which is also ultimately used for KIE.
This strategy can be helpful in real-world settings, in which a detailed hierarchical output of input documents is required, such as an adequate differentiation between individual line items in invoices. Non-generative KIE methods that for example decode outputs with a sequence labeling layer usually have difficulties in reconstructing such hierarchies and require additional post-processing steps.
To let the model learn text recognition, \cite{nr59, nr75} use a pre-training task where the model must predict the next token while considering the previous tokens as well as the document image.

\subsection{Practical perspective}
\label{domain}

Based on the analysis, there is a clear lack of a practice-oriented perspective that is being adopted in related work. To this end, only a small number of the analyzed approaches integrate domain knowledge. Moreover, in most of these cases, domain knowledge is at most fused into the systems by the means of hand-crafted input features. There is also a lack of evaluating proposed KIE approaches and their impact on real-world scenarios. These observations have also been made in the study by \cite{R6}.

One work that stands out in terms of its consideration of domain knowledge is the approach by \cite{nr51}. The authors make use of such knowledge by proposing a hybrid KIE system based on both DL-models and rule-based methods including several post-processing steps to improve the extraction results. This includes, for example, the correction of automatically extracted product codes and product prices. However, this post-processing is considered as a supplement to the DL model in case of incorrect predictions and not an architecture-based adaptation.
Another example is the work by \cite{nr8}, which integrates domain-oriented constraints regarding invoices that the total amount is the sum of the subtotal and tax total values, and that the tax total can be computed as the product of subtotal and tax percentage. The authors achieve this by adjusting the loss calculations during training, but report that no significant performance improvements were achieved. At the same time, this emphasizes the need for further research in this area (see also chapter \ref{agenda}). 
While not directly implementing such aspects in the proposed methods, some authors also designed their approaches in a way to allow for integrating domain knowledge. For example, in the work of \cite{nr46}, candidates are first identified for fields to be extracted and subsequently scored. In this regard, one example the authors mention is to define a scoring function that integrates constraints such as the fact that an invoice date must precede its due date.

As discussed in section \ref{kie_processes}, adopting a practical perspective and considering domain knowledge can provide valuable insights for an automated extraction system in various ways. However, this aspect has not been properly explored in the literature, even though it has been named as a key challenge for DU in the past \citep{docintelligence}. Therefore, future work should consider adopting a business process perspective in order to develop approaches that exploit this unrealized potential.
This could also lead to the identification of completely new potentials of KIE systems that are not yet considered in current research, as the current focus often lies only on increasing the performance with respect to established benchmark datasets.
These considerations may require a deeper understanding of the associated business processes, which can be achieved by modeling computer-integrated systems including data, behavior and control flows \citep{heraklit}.
The potential of context-aware DL systems has been investigated in other research areas. For example, \cite{ie_ppm} propose the fusion of information extracted from business documents with traditional event log data in order to enhance predictive process monitoring capabilities. However, the other direction, i.e., feeding context-aware data obtained from process mining techniques into KIE systems, has not yet been considered.

\subsection{Trends}

One can observe several trends in KIE research since 2017 in a wide variety of facets. 
One aspect is the dominance of certain paradigms of KIE systems -- most notably sequence-based methods, which represent the predominant approach since 2021. To the contrary, grid-based methods have never been able to establish themselves, despite their comparability with graph-based approaches, which have been popular since the beginning. At the same time, this indicates that VRDs with usually complex layouts cannot be appropriately represented as grids with well-defined structures that are less flexible compared to dynamic graphs. As expected, LLM-based systems have emerged in 2023 and quickly became popular.
Also, using conjunctions of different paradigms (e.g., sequence-based and graph-based) is not as widely adopted as the focus on one particular paradigm.

Since 2022, an increased focus has been placed on creating OCR-independent and generative KIE approaches. The latter is the consequence of the increasing popularity of generative AI methods such as LLMs in recent years and their influence on the DU domain. The increasing shift towards OCR-independent systems can be explained by the otherwise necessary external OCR engines, which can be error-prone and therefore lead to falsely extracted information, especially in real-world use cases \cite{R3}.

It can also be observed that over time, there have been different strategies on how to develop and improve KIE systems. While towards the beginning, a lot of research effort has been placed on how to properly integrate the different input modalities, other strategies have also emerged in recent years. For example, some publications highlight and investigate the importance of the reading order. In traditional methods, also as a consequence of the reliance on external OCR engines, a simple top-bottom and left-right reading order is usually adopted. This however can lead to a suboptimal segmentation of complex VRDs. To this end, approaches such as those by \cite{nr63, nr88, nr94} investigate more sophisticated methods to obtain an optimized reading order that better suits the actual document layout. \cite{metrics_for_order} also propose evaluation metrics that overcome biases resulting from the reading order.
However, what was not or hardly ever considered as a lever for better KIE performance, is the model size in terms of trainable parameters. Between 2017 and 2024, there was no significant increase in model sizes besides a more frequent use of LLMs.

It is also positive to note that code and/or model weights are being published more frequently in recent years. Even if the absolute number of implementations that have been shared is relatively low, it has become increasingly more frequent, especially since 2021. This is a positive development, as it can accelerate research progress and research dissemination. This observation also goes hand in hand with the increased popularity of HuggingFace\footnote{\url{https://huggingface.co/}}, which is a platform that provides tools, datasets, and pre-trained models to facilitate research in NLP and CV. Some previously discussed KIE approaches are also available, for example LayoutLMv3\footnote{\url{https://huggingface.co/microsoft/layoutlmv3-base}}.

It can be assumed that these trends and observations will continue in future KIE research. Especially a shift towards generative KIE systems seems evident, as corresponding models become more and more popular across multiple domains. Initial approaches that make use of powerful LLMs exist, however they do not consistently achieve competitive results compared to specialized DU methods yet. 
In this regard, additional research is required on how to close this performance gap and thus make corresponding approaches more viable, especially with regard to the trade-off between model size (and therefore hardware requirements) and extraction performance as discussed in section \ref{quantitative_comparison}.

\section{Research agenda}
\label{agenda}

We have identified several aspects that should be considered in future work, which could not only lead to better research results, but also to a better applicability of KIE systems in real-world scenarios.

{\textbf{\textit{Novel datasets.}}} Only a small number of public datasets are commonly used. For example, more than half of the sequence-based approaches are pre-trained using (subsets of) IIT-CDIP. One problem with this dataset in particular is the relatively low image quality, which no longer meets today's standards. These properties have a direct impact on the use of corresponding approaches in real-world settings, which typically involve documents with better image quality.

In addition to pre-training datasets, efforts should also be made to construct novel benchmark datasets, ideally based on different document types compared to commonly considered receipts -- as also discussed in \cite{R7, bd_benchmarks}. Besides different domains, newly created datasets should also be designed to more closely resemble documents found in real-world scenarios. \cite{nr92, nr94, eval_tool} have shown that existing dominant datasets have numerous shortcomings in this respect. Also, some datasets contain erroneous annotations and inconsistent labeling behavior \cite{nr112}.
Furthermore, common benchmarks such as SROIE and FUNSD show a high degree of layout replication in training and test partitions \cite{generalizability}, which distorts the overall validity the reported evaluation results. 
One possible solution to construct novel datasets can be the generation of synthetic documents, however, there is a risk that synthetic documents do not have the properties of real-world documents in terms of layout variety and complexity \citep{bd_benchmarks}.

{\textbf{\textit{Consistent evaluation.}}} The analysis in section \ref{eval} has shown that the papers are very heterogeneous in terms of their evaluation setup. In addition, the authors often do not specify their evaluation methodology in much detail, which raises questions on how exactly the evaluation results were obtained. Therefore, a quantitative comparison of the approaches is not always meaningful. This is especially problematic since the understanding of improving the state of the art in this research area is often associated with an increase in performance on benchmark datasets compared to existing work.
Some public competitions, such as SROIE, provide a dedicated evaluation protocol for a consistent ranking of methods. \footnote{The leaderboard for the KIE task can be found at \url{https://rrc.cvc.uab.es/?ch=13&com=evaluation&task=3}.} 
However, this is the exception rather than the rule.
    
Therefore, it would be desirable to agree on a consistent approach or implement a centralized evaluation toolkit that could be used for different benchmark datasets, where a specific set of metrics (e.g., Precision, Recall, F1) is implemented. \cite{DUE} have proposed a reference implementation for a DU benchmark, but it is not yet widely used.
We also advocate a more frequent use of string-based evaluation metrics, as they allow for a better assessment in real-world applications. However, only one third of the analyzed manuscripts used such string-based evaluation setups.
It is also surprising that only about a third of the papers present the results on field-level. In this regard, we advocate that such an evaluation should be presented more frequently in future work, as it gives a good indication of whether an approach faces problems with certain fields, which in turn allows for a more in-depth analysis, as also discussed by \cite{new_agenda}.
Last, but not least, authors should adopt entity-level performance measures, as it is more suitable to assess the performance for practical applications where the extracted entities are processed as a whole \citep{nr94}. The need for performance measures that provide a better indication of real-world performance is also addressed by \cite{new_metric}, which propose a novel metric that shall mitigate the aforementioned shortcomings.

{\textbf{\textit{Pre-processing methods.}}} There exists a dedicated research area that investigates techniques for pre-processing document images. An example of a sophisticated approach is proposed by \cite{preprocessing_example}. In this context, synergy effects with this research area should be exploited, in particular the integration of corresponding techniques into KIE systems \cite{R2}. Related work often does not include pre-processing techniques, or, if they are used, only simple methods such as Deskewing.
During the life cycle of a document in real-world settings, there are various points at which image quality can be impaired, for example due to introduced scan artifacts by digitization steps or as a consequence of multiple compression and transmission procedures \citep{diqa}. \cite{aesthetics} show, that quality aspects such as image noise or fonts can have significant impacts on the performance of DU systems. Therefore, more emphasis should be put on the adoption of respective methods in DU pipelines.

{\textbf{\textit{Tokenization.}}} In particular sequence-based KIE approaches utilize tokenizers for the encoding of input documents into token sequences, which are subsequently supplemented by visual and positional embeddings. Whenever the KIE systems incorporate pre-trained language models such as BERT as their encoder backbone, they usually also adopt the corresponding tokenizer without further adjustments. 
However, there have been studies highlighting the challenges associated with the adoption of unaltered tokenizers to different domains \citep{bert_tokenizer_challenges}. To this end, several methods have been proposed to adjust tokenizers to new domains, which improves the performance on downstream tasks. Examples are domain-specific augmentations of the original tokenizers' vocabulary \citep{exbert,adaptive_tokenization}, or a careful investigation of training data, pre-tokenization setups and other changes to the vocabulary \citep{tokenizer_adaptions}.
    
Currently, there is no extensive investigation of the role of tokenizers in the context of KIE. \cite{tokenization_ner} analyze the impact of tokenization regarding NER from biomedical texts, however there is a lack of research considering complex DU tasks on VRDs.
Therefore, future work should consider exploring sophisticated methods to allow for an adequate transfer of existing tokenizers to KIE and/or develop novel methods for tokenizing complex documents in a more robust manner. 
Conceivable first steps could be to initially use existing KIE approaches based on out-of-the-box tokenizers, only focus on the adaptation of the tokenizer and perform benchmarks against the original implementation.

{\textbf{\textit{Generalizability.}}} \cite{generalizability_tested} have shown that existing KIE methods lack adequate generalization capabilities, which may be due to the limitations associated with the datasets as mentioned before. Furthermore, the aspect of generalizability is often not considered in detail. Instead, it is usually only implicitly considered when the approaches are evaluated based on multiple benchmark datasets with different document types.

Future work should focus on how to effectively design model architectures as well as novel (pre-)training tasks in order to achieve a higher degree of generalizability across different document layouts, types and domains. 
Another possibility is to investigate the modularization of KIE systems more closely in order to reduce interdependencies between individual components and thus obtain more generic approaches \citep{nr8}.
There are some efforts in other domains such as dental image analysis \citep{generalizability_images}. Research has also been conducted in the context of DLA, where the authors advocate the combination of novel models and curated (synthetic) datasets to address the challenge of generalizability \citep{generalizability_dla}. However, there are not many efforts in the context of KIE.

{\textbf{\textit{Domain knowledge.}}} As discussed in chapter \ref{domain}, only a few authors integrate domain knowledge into the proposed approaches or adopt a practice-oriented view in general. However, it seems promising to integrate existing domain knowledge into KIE systems, which is also discussed in \cite{R3}, as it contains valuable information about the corresponding business processes, their documents and how they need to be understood as a whole. The design of novel pre-training tasks could be one possibility for integrating these aspects.

{\textbf{\textit{Real-world usability.}}} The relevance of automated document processing in real-world scenarios is indisputable (see also the review by \cite{R6}). Because of this great importance, more efforts should be made to improve the overall practicability of KIE approaches. 
The previously mentioned research directions already address this matter. For example, one could consider pre-processing techniques to remove document image artifacts that are specific to real-world settings, such as removing stamps \citep{stamp}.  
The aforementioned focus on generalization capabilities can also have a positive impact on the practicality of KIE systems since deploying a system that can properly process many different types of documents simultaneously could result in lower costs and less maintenance.
    
On top of that, additional aspects could be considered. For example, \cite{explain_preds} propose a method to provide confidence estimates for the extracted data. The authors emphasize that in industrial settings, the primary goal is typically to make decisions based on model predictions rather than the raw extracted data. Research in this direction could greatly assist the collaboration between KIE systems and human administrators during document processing tasks, as it is essential to validate automatically extracted data in real-world settings \citep{review}. 
\cite{R8} also emphasize the importance of real-time KIE in industry applications. In this regard, a focus on lightweight KIE solutions should be considered. This could for example include investigating the trade-off between model size and extraction performance more closely. \cite{nr80} have shown that a relatively small model with around 50 thousand parameters can produce competitive results compared to systems that consist of several hundred million parameters.

{\textbf{\textit{End-to-end performance.}}} The evaluation has shown that end-to-end approaches usually achieve inferior extraction results compared to systems that use an external OCR engine. 
In this regard, more research should be conducted on how to improve the text recognition step, as the end-to-end systems have a high potential due to their independence of OCR engines, also highlighted by \cite{R8}. 
End-to-end approaches have received less attention so far, which is also reflected in the number of identified approaches. Nevertheless, there is a relatively strong increase in corresponding methods, especially since 2022.
\section{Conclusion}
\label{conclusion}

Research in the area of Key Information Extraction has seen an increased interest in recent years, mainly due to major advances in the field of Deep Learning. Nowadays, even visually-rich business documents with very complex layouts and high information-richness can be automatically processed by corresponding systems. 
To this end, this manuscript represents a systematic literature review covering the research on Key Information Extraction between 2017 and 2024 including 130 proposed approaches, with the aim of identifying the state of the art in this field and identifying potentials for further research. The identified methods were compared both qualitatively and quantitatively based on various characteristics.

The analysis has shown that related work tends to follow a very narrow corridor in which already proven concepts are being successively refined and improved. In general, the approaches follow the same paradigms for representing document images, namely as sequences, graphs or grids. In detail, the approaches differ in their choice of architectures used for encoding and decoding the input documents, although it can also be seen that certain models are used more frequently (e.g., BERT-based models for textual inputs).
Besides, novel concepts have been explored over time, such as OCR-independent and autoregressive methods, which on the one hand do not require an external OCR engine for preliminary text reading steps, and on the other hand can output arbitrary text, making them more flexible in terms of the downstream tasks they support.
The authors investigate how different input modalities implicitly and explicitly contained in complex documents can be optimally integrated into the model architectures. In particular, visual cues from document images are increasingly incorporated into the models in order to improve their performance for more complex use cases.
Much effort is also put into learning a general document understanding through Deep Learning models, which can then be used for Key Information Extraction. This is usually done by extensive and innovative pre-training tasks followed by specific fine-tuning steps.
Another general observation is that the complexity of the corresponding models in terms of the number of parameters does not play a significant role and that even lightweight models can achieve promising extraction results.

The research area is strongly characterized by the fact that an improvement of the state of the art is achieved by obtaining better extraction results on established benchmark datasets. However, a quantitative comparison of the presented results is not always meaningful, since the evaluation setups used by the authors are very heterogeneous. Furthermore, for most of the benchmarks, very good results have already been achieved (F1-scores above 0.97). The research area should therefore move away from focusing on improving benchmarks results and instead investigate innovations along other dimensions. 
Future work could focus on even more lightweight models in order to improve their practical applicability. It could also be investigated how the models can be designed in order to require less data for training procedures. We identified several other starting points for follow-up research based on the findings. These include proposing novel and more diverse datasets as well as consistent evaluation setups that allow for an adequate quantitative comparison. We also advocate that more attention should be paid to the real-world usability of corresponding approaches. This includes the integration of domain knowledge, since on the one hand document processing tasks play a key role in daily business workloads and, on the other hand, offer a special perspective on Key Information Extraction with unique requirements, but also possibilities.

\bibliographystyle{ACM-Reference-Format}
\bibliography{bibliography}

\newpage

\section{Supplementary material}

\setcounter{section}{0}
\setcounter{table}{0}
\setcounter{figure}{0}

\renewcommand\thesection{\Alph{section}}
\renewcommand{\thetable}{ST\arabic{table}}
\renewcommand{\thefigure}{SF\arabic{figure}}

\section{Business process perspective}

\begin{figure}[!ht]
    \centering
    \caption{Exemplary run of a purchasing process}
    \label{process_run}
    \Description{A graph that shows a run of a purchasing process with document-based exchanges highlighted.}
    \rotatebox{90}{\includegraphics[width=0.87\textheight]{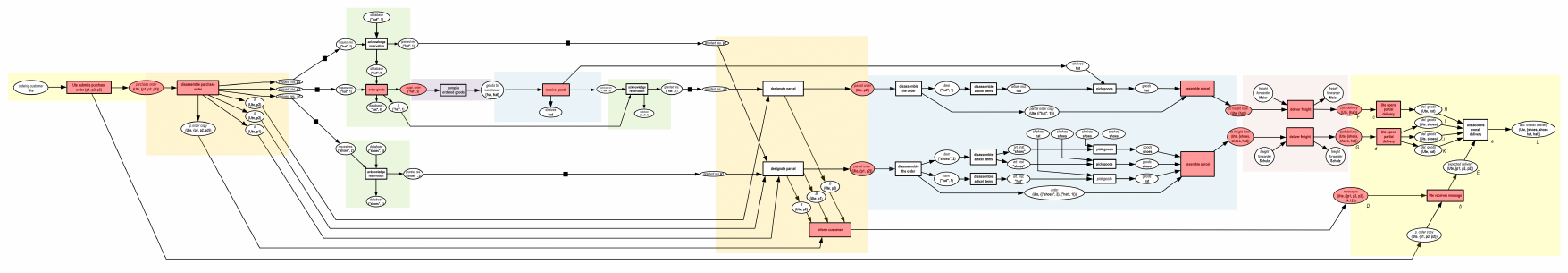}}
\end{figure}
\newpage
\section{Details for search procedure}

\subsection{Database search strings}

\begin{table}[!ht]
\caption{Defined search strings}
\label{strings}
\begin{tabular}{p{0.06\textwidth}p{0.82\textwidth}>{\centering\arraybackslash}p{0.05\textwidth}}
\toprule
\textbf{DB} & \textbf{Search string} & \textbf{Hits} \\ 
\midrule
ACL & intitle:((information AND extract*) OR "entity extraction" OR "entity recognition" OR document* OR "visually rich") AND (intitle:("visually?rich" OR modal* OR image* OR invoice* OR receipt*) OR abstract:("visually?rich" OR modal* OR image* OR invoice* OR receipt*) OR keywords:("visually?rich" OR modal* OR image* OR invoice* OR receipt*)) & 100 \\ \midrule
ACM & Title:((information AND extract*) OR "entity extraction" OR "entity recognition" OR document* OR "visually?rich") AND ( Title:("visually?rich" OR modal* OR image OR invoice OR receipt) OR Abstract:("visually?rich" OR modal* OR image OR invoice OR receipt) OR Keyword:("visually?rich" OR modal* OR image OR invoice OR receipt) ) AND NOT (Title:(handwrit*) OR Title:(histor*) OR Title:(table) OR Title:(web)) & 300 \\ \midrule
AIS & ((title:information AND title:extract*) OR title:"entity extraction" OR title:"entity recognition" OR title:document* OR title:"visually rich") AND ("deep learning" OR "artificial intelligence" OR "neural network*") \&start\_date=01/01/2017\&end\_date=12/31/2023 & 35 \\ \midrule
IEEE & ((("Document Title":information AND "Document Title":extract*) OR "Document Title":"entity extraction" OR "Document Title":"entity recognition" OR "Document Title":document* OR "Document Title":"visually rich") AND ( ("Document Title":"visually rich" OR "Document Title":modal* OR "Document Title":invoice OR "Document Title":receipt) OR ("Abstract":"visually rich" OR "Abstract":modal* OR "Abstract":invoice OR "Abstract":receipt) OR ("Author Keywords":"visually rich" OR "Author Keywords":modal* OR "Author Keywords":invoice OR "Author Keywords":receipt) ) NOT ("Document Title":handwrit* OR "Document Title":histor* OR "Document Title":table)) & 152 \\ \midrule
SD & Year: 2017-2023; Title, abstract, keywords: "visually rich" OR modal OR image OR invoice OR receipt; Title: (information AND extract) OR document OR "visually rich" ; "deep learning" OR "artificial intelligence" OR cognitive OR intelligent OR "neural network" & 198 \\ \midrule
SL & query="visually+rich"+OR+modal*+OR+image+OR+invoice+OR+receipt \& dc.title=information+extract* \& date-facet-mode=between\&facet-start-year=2017 \& facet-end-year=2023 & 348 \\ \midrule
\textbf{Total} &  & \textbf{1133} \\ 
\bottomrule
\end{tabular}
\end{table}

The title should include the name of the DU subtask, i.e., KIE, its variant formulations, and other commonly used terms such as entity extraction. We also included \textit{document*} in the title search. This in particular resulted in many non-relevant papers, but is important, as there are many other terms used for this research area, as mentioned before. This increased the manual effort for filtering, but also increased the recall and thus the coverage.
Furthermore, titles, abstracts, keywords or metadata should include terms such as \textit{visually rich} or \textit{image} in order to find literature that specifically deals with VRDs according to the scope of this work. We also include terms such as \textit{invoice} or \textit{receipts}, since these are the most common terms regarding document types. However, these terms were included with OR-operators, which means that the results do not necessarily have to include them. 
To identify approaches using DL methods, we added the terms \textit{deep learning}, \textit{artificial intelligence} and \textit{neural network*}, where at least one of them should appear somewhere in the text or metadata.
In order to filter out papers that are outside the scope of this work, we included terms that should not appear in the title, namely \textit{handwrit*}, \textit{histor*}, \textit{table} or \textit{web}, if supported by the search engine.
We also used wildcards (*) to allow for different expressions of the search terms (e.g., \textit{(information AND extract*)}). 

\subsection{Prisma diagram}

\begin{figure}[!ht]
    \centering
    \includegraphics[width=0.8\textwidth]{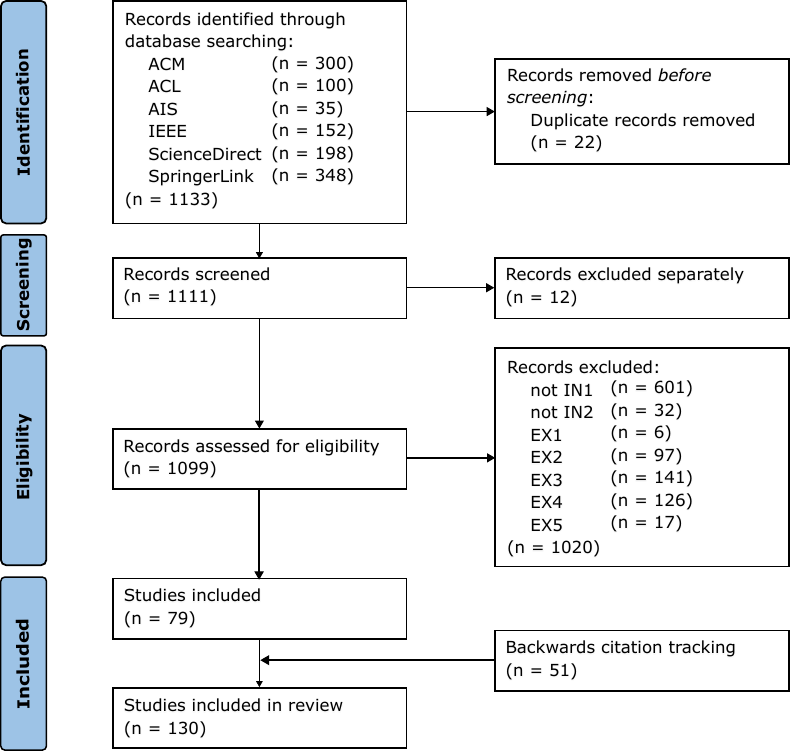}
    \caption{PRISMA flow diagram}
    \Description{Literature search procedure according to PRISMA flow diagram}
    \label{prisma}
\end{figure} 
\newpage
\section{Numbering of the approaches}
The manuscripts are sorted chronologically by year of publication. Within a year, they are sorted alphabetically by author names.

\begin{longtable}{>{\centering\arraybackslash}p{0.03\textwidth} p{0.1\textwidth} p{0.8\textwidth}}
\caption{Numbering of the approaches}
\label{mapping} \\
\toprule
\textbf{\#} & \textbf{Reference} & \textbf{Title} \\
\midrule
1 & \cite{nr1} \citeyear{nr1} & CloudScan - A Configuration-Free Invoice Analysis System Using Recurrent Neural Networks \\ \hline
2 & \cite{nr2} \citeyear{nr2} & Chargrid: Towards understanding 2D documents \\ \hline
3 & \cite{nr3} \citeyear{nr3} & BERTgrid: Contextualized Embedding for 2D Document Representation and Understanding \\ \hline
4 & \cite{nr4} \citeyear{nr4} & Visual-Linguistic Methods for Receipt Field Recognition \\ \hline
5 & \cite{nr5} \citeyear{nr5} & EATEN: Entity-aware attention for single shot visual text extraction \\ \hline
6 & \cite{nr6} \citeyear{nr6} & Graph convolution for multimodal information extraction from visually rich documents \\ \hline
7 & \cite{nr7} \citeyear{nr7} & An Invoice Reading System Using a Graph Convolutional Network \\ \hline
8 & \cite{nr8} \citeyear{nr8} & Attend, copy, parse end-to-end information extraction from documents \\ \hline
9 & \cite{nr9} \citeyear{nr9} & Graphie: A graph-based framework for information extraction \\ \hline
10 & \cite{nr10} \citeyear{nr10} & Named entity recognition and relation extraction with graph neural networks in semi structured documents \\ \hline
11 & \cite{nr11} \citeyear{nr11} & One-shot Text Field labeling using Attention and Belief Propagation for Structure Information Extraction \\ \hline
12 & \cite{nr12} \citeyear{nr12} & End-to-end information extraction by character-level embedding and multi-stage attentional u-net \\ \hline
13 & \cite{nr13} \citeyear{nr13} & Cardinal Graph Convolution Framework for Document Information Extraction \\ \hline
14 & \cite{nr14} \citeyear{nr14} & Attention-Based Graph Neural Network with Global Context Awareness for Document Understanding \\ \hline
15 & \cite{nr15} \citeyear{nr15} & Merge and Recognize: A Geometry and 2D Context Aware Graph Model for Named Entity Recognition from Visual Documents \\ \hline
16 & \cite{nr16} \citeyear{nr16} & Representation learning for information extraction from form-like documents \\ \hline
17 & \cite{nr17} \citeyear{nr17} & Information Extraction from Text Intensive and Visually Rich Banking Documents \\ \hline
18 & \cite{nr18} \citeyear{nr18} & End-to-End Extraction of Structured Information from Business Documents with Pointer-Generator Networks \\ \hline
19 & \cite{nr19} \citeyear{nr19} & DocStruct: A multimodal method to extract hierarchy structure in document for general form understanding \\ \hline
20 & \cite{nr20} \citeyear{nr20} & Robust Layout-aware IE for Visually Rich Documents with Pre-trained Language Models \\ \hline
21 & \cite{nr21} \citeyear{nr21} & LayoutLM: Pre-training of Text and Layout for Document Image Understanding \\ \hline
22 & \cite{nr22} \citeyear{nr22} & Pick: Processing key information extraction from documents using improved graph learning-convolutional networks \\ \hline
23 & \cite{nr23} \citeyear{nr23} & TRIE: End-to-End Text Reading and Information Extraction for Document Understanding \\ \hline
24 & \cite{nr24} \citeyear{nr24} & DocFormer: End-to-End Transformer for Document Understanding \\ \hline
25 & \cite{nr25} \citeyear{nr25} & Consideration of the Word's Neighborhood in GATs for Information Extraction in Semi-structured Documents \\ \hline
26 & \cite{nr26} \citeyear{nr26} & Visual FUDGE: Form Understanding via Dynamic Graph Editing \\ \hline
27 & \cite{nr27} \citeyear{nr27} & LAMBERT: Layout-Aware Language Modeling for Information Extraction \\ \hline
28 & \cite{nr28} \citeyear{nr28} & Unified Pretraining Framework for Document Understanding \\ \hline
29 & \cite{nr29} \citeyear{nr29} & Information Extraction from Invoices \\ \hline
30 & \cite{nr30} \citeyear{nr30} & Learning from similarity and information extraction from structured documents \\ \hline
31 & \cite{nr31} \citeyear{nr31} & Cost-effective End-to-end Information Extraction for Semi-structured Document Images \\ \hline
32 & \cite{nr32} \citeyear{nr32} & Spatial Dependency Parsing for Semi-Structured Document Information Extraction \\ \hline
33 & \cite{nr33} \citeyear{nr33} & VisualWordGrid: Information Extraction from Scanned Documents Using a Multimodal Approach \\ \hline
34 & \cite{nr34} \citeyear{nr34} & DocReader: Bounding-Box Free Training of a Document Information Extraction Model \\ \hline
35 & \cite{nr35} \citeyear{nr35} & Information Extraction from Invoices: A Graph Neural Network Approach for Datasets with High Layout Variety \\ \hline
36 & \cite{nr36} \citeyear{nr36} & ROPE: Reading Order Equivariant Positional Encoding for Graph-based Document Information Extraction \\ \hline
37 & \cite{nr37} \citeyear{nr37} & StructuralLM: Structural pre-training for form understanding \\ \hline
38 & \cite{nr38} \citeyear{nr38} & SelfDoc: Self-Supervised Document Representation Learning \\ \hline
39 & \cite{nr39} \citeyear{nr39} & StrucTexT: Structured Text Understanding with Multi-Modal Transformers \\ \hline
40 & \cite{nr40} \citeyear{nr40} & ViBERTgrid: A Jointly Trained Multi-modal 2D Document Representation for Key Information Extraction from Documents \\ \hline
41 & \cite{nr41} \citeyear{nr41} & A Span Extraction Approach for Information Extraction on Visually-Rich Documents \\ \hline
42 & \cite{nr42} \citeyear{nr42} & A Segment-Based Layout Aware Model for Information Extraction on Document Images \\ \hline
43 & \cite{nr43} \citeyear{nr43} & Going Full-TILT Boogie on Document Understanding with Text-Image-Layout Transformer \\ \hline
44 & \cite{nr44} \citeyear{nr44} & Improving information extraction from visually rich documents using visual span representations \\ \hline
45 & \cite{nr45} \citeyear{nr45} & MatchVIE: Exploiting Match Relevancy between Entities for Visual Information Extraction \\ \hline
46 & \cite{nr46} \citeyear{nr46} & Glean: Structured extractions from templatic documents \\ \hline
47 & \cite{nr47} \citeyear{nr47} & Towards Robust Visual Information Extraction in Real World: New Dataset and Novel Solution \\ \hline
48 & \cite{nr48} \citeyear{nr48} & Tag, Copy or Predict: A Unified Weakly-Supervised Learning Framework for Visual Information Extraction using Sequences \\ \hline
49 & \cite{nr49} \citeyear{nr49} & LayoutLMv2: Multi-modal pre-training for visually-rich document understanding \\ \hline
50 & \cite{nr50} \citeyear{nr50} & Entity Relation Extraction as Dependency Parsing in Visually Rich Documents \\ \hline
51 & \cite{nr51} \citeyear{nr51} & Key Information Extraction in Purchase Documents using Deep Learning and Rule-based Corrections \\ \hline
52 & \cite{nr52} \citeyear{nr52} & Query-driven Generative Network for Document Information Extraction in the Wild \\ \hline
53 & \cite{nr53} \citeyear{nr53} & GMN: Generative Multi-modal Network for Practical Document Information Extraction \\ \hline
54 & \cite{nr54} \citeyear{nr54} & Automatic Key Information Extraction from Visually Rich Documents \\ \hline
55 & \cite{nr55} \citeyear{nr55} & CALM: Commen-Sense Knowledge Augmentation for Document Image Understanding \\ \hline
56 & \cite{nr56} \citeyear{nr56} & XYLayoutLM: Towards Layout-Aware Multimodal Networks For Visually-Rich Document Understanding \\ \hline
57 & \cite{nr57} \citeyear{nr57} & BROS: A Pre-trained Language Model Focusing on Text and Layout for Better Key Information Extraction from Documents \\ \hline
58 & \cite{nr58} \citeyear{nr58} & LayoutLMv3: Pre-training for Document AI with Unified Text and Image Masking \\ \hline
59 & \cite{nr59} \citeyear{nr59} & OCR-Free Document Understanding Transformer \\ \hline
60 & \cite{nr60} \citeyear{nr60} & FormNet: Structural Encoding beyond Sequential Modeling in Form Document Information Extraction \\ \hline
61 & \cite{nr61} \citeyear{nr61} & Relational Representation Learning in Visually-Rich Documents \\ \hline
62 & \cite{nr62} \citeyear{nr62} & Fusion of visual representations for multimodal information extraction from unstructured transactional documents \\ \hline
63 & \cite{nr63} \citeyear{nr63} & ERNIE-Layout: Layout Knowledge Enhanced Pre-training for Visually-rich Document Understanding \\ \hline
64 & \cite{nr64} \citeyear{nr64} & LiLT: A Simple yet Effective Language-Independent Layout Transformer for Structured Document Understanding \\ \hline
65 & \cite{nr65} \citeyear{nr65} & MmLayout: Multi-grained MultiModal Transformer for Document Understanding \\ \hline
66 & \cite{nr66} \citeyear{nr66} & Tokengrid: Toward More Efficient Data Extraction From Unstructured Documents \\ \hline
67 & \cite{nr67} \citeyear{nr67} & Extract Data Points from Invoices with Multi-layer Graph Attention Network and Named Entity Recognition \\ \hline
68 & \cite{nr68} \citeyear{nr68} & Dual-VIE: Dual-Level Graph Attention Network for Visual Information Extraction \\ \hline
69 & \cite{nr69} \citeyear{nr69} & Low-Dimensionality Information Extraction Model for Semi-structured Documents \\ \hline
70 & \cite{nr70} \citeyear{nr70} & Improving Information Extraction from Semi-structured Documents Using Attention Based Semi-variational Graph Auto-Encoder \\ \hline
71 & \cite{nr71} \citeyear{nr71} & GenKIE: Robust Generative Multimodal Document Key Information Extraction \\ \hline
72 & \cite{nr72} \citeyear{nr72} & End-to-End Document Recognition and Understanding with Dessurt \\ \hline
73 & \cite{nr73} \citeyear{nr73} & An Iterative Graph Learning Convolution Network for Key Information Extraction Based on the Document Inductive Bias \\ \hline
74 & \cite{nr74} \citeyear{nr74} & GenTC: Generative Transformer via Contrastive Learning for Receipt Information Extraction \\ \hline
75 & \cite{nr75} \citeyear{nr75} & DocParser: End-to-end OCR-Free Information Extraction from Visually Rich Documents \\ \hline
76 & \cite{nr76} \citeyear{nr76} & A Novel Approach for Extracting Key Information from Vietnamese Prescription Images \\ \hline
77 & \cite{nr77} \citeyear{nr77} & VisualIE: Receipt-Based Information Extraction with a Novel Visual and Textual Approach \\ \hline
78 & \cite{nr78} \citeyear{nr78} & Doc2Graph: A Task Agnostic Document Understanding Framework Based on Graph Neural Networks \\ \hline
79 & \cite{nr79} \citeyear{nr79} & Enhancing GNN Feature Modeling for Document Information Extraction Using Transformers \\ \hline
80 & \cite{nr80} \citeyear{nr80} & ICL-D3IE: In-Context Learning with Diverse Demonstrations Updating for Document Information Extraction \\ \hline
81 & \cite{nr81} \citeyear{nr81} & Document Information Extraction via Global Tagging \\ \hline
82 & \cite{nr82} \citeyear{nr82} & Visual Information Extraction in the Wild: Practical Dataset and End-to-End Solution \\ \hline
83 & \cite{nr83} \citeyear{nr83} & FormNetV2: Multimodal Graph Contrastive Learning for Form Document Information Extraction \\ \hline
84 & \cite{nr84} \citeyear{nr84} & Enhancing Visually-Rich Document Understanding via Layout Structure Modeling \\ \hline
85 & \cite{nr85} \citeyear{nr85} & DocTr: Document Transformer for Structured Information Extraction in Documents \\ \hline
86 & \cite{nr86} \citeyear{nr86} & GeoLayoutLM: Geometric Pre-training for Visual Information Extraction \\ \hline
87 & \cite{nr87} \citeyear{nr87} & Multi-scale Cell-based Layout Representation for Document Understanding \\ \hline
88 & \cite{nr88} \citeyear{nr88} & Unifying Vision, Text, and Layout for Universal Document Processing \\ \hline
89 & \cite{nr89} \citeyear{nr89} & LayoutMask: Enhance Text-Layout Interaction in Multi-modal Pre-training for Document Understanding \\ \hline
90 & \cite{nr90} \citeyear{nr90} & DocGraphLM: Documental Graph Language Model for Information Extraction \\ \hline
91 & \cite{nr91} \citeyear{nr91} & Modeling Entities as Semantic Points for Visual Information Extraction in the Wild \\ \hline
92 & \cite{nr92} \citeyear{nr92} & UReader: Universal OCR-free Visually-situated Language Understanding with Multimodal Large Language Model \\ \hline
93 & \cite{nr93} \citeyear{nr93} & StrucTexTv2: Masked Visual-Textual Prediction for Document Image Pre-training \\ \hline
94 & \cite{nr94} \citeyear{nr94} & Reading Order Matters: Information Extraction from Visually-rich Documents by Token Path Prediction \\ \hline
95 & \cite{nr95} \citeyear{nr95} & A Character-Level Document Key Information Extraction Method with Contrastive Learning \\ \hline
96 & \cite{nr96} \citeyear{nr96} & Multimodal Pre-training Based on Graph Attention Network for Document Understanding \\ \hline
97 & \cite{nr97} \citeyear{nr97} & Enhancing Document Information Analysis with Multi-Task Pre-training: A Robust Approach for Information Extraction in Visually-Rich Documents \\ \hline
98 & \cite{nr98} \citeyear{nr98} & AIE-KB: Information Extraction Technology with Knowledge Base for Chinese Archival Scenario \\ \hline
99 & \cite{nr99} \citeyear{nr99} & GeoContrastNet: Contrastive Key-Value Edge Learning for Language-Agnostic Document Understanding \\ \hline
100 & \cite{nr100} \citeyear{nr100} & GraphMLLM: A Graph-Based Multi-level Layout Language-Independent Model for Document Understanding \\ \hline
101 & \cite{nr101} \citeyear{nr101} & DocPedia: unleashing the power of large multimodal model in the frequency domain for versatile document understanding \\ \hline
102 & \cite{nr102} \citeyear{nr102} & LayoutLLM: Large Language Model Instruction Tuning for Visually Rich Document Understanding \\ \hline
103 & \cite{nr103} \citeyear{nr103} & Information Extraction from Visually Rich Documents Using Directed Weighted Graph Neural Network \\ \hline
104 & \cite{nr104} \citeyear{nr104} & Multimodal weighted graph representation for information extraction from visually rich documents \\ \hline
105 & \cite{nr105} \citeyear{nr105} & DCMAI: A Dynamical Cross-Modal Alignment Interaction Framework for Document Key Information Extraction \\ \hline
106 & \cite{nr106} \citeyear{nr106} & mPLUG-DocOwl 1.5: Unified Structure Learning for OCR-free Document Understanding \\ \hline
107 & \cite{nr107} \citeyear{nr107} & UniVIE: A Unified Label Space Approach to Visual Information Extraction from Form-Like Documents \\ \hline
108 & \cite{nr108} \citeyear{nr108} & MCKIE: Multi-class Key Information Extraction from Complex Documents Based on Graph Convolutional Network \\ \hline
109 & \cite{nr109} \citeyear{nr109} & Reformulating Key-Information Extraction as Next Sentence Prediction for Hierarchical Data \\ \hline
110 & \cite{nr110} \citeyear{nr110} & ProtoNER: Few Shot Incremental Learning for Named Entity Recognition Using Prototypical Networks \\ \hline
111 & \cite{nr111} \citeyear{nr111} & DocPointer: A parameter-efficient Pointer Network for Key Information Extraction \\ \hline
112 & \cite{nr112} \citeyear{nr112} & PEneo: Unifying Line Extraction, Line Grouping, and Entity Linking for End-to-end Document Pair Extraction \\ \hline
113 & \cite{nr113} \citeyear{nr113} & HRVDA: High-Resolution Visual Document Assistant \\ \hline
114 & \cite{nr114} \citeyear{nr114} & LayoutLLM: Layout Instruction Tuning with Large Language Models for Document Understanding \\ \hline
115 & \cite{nr115} \citeyear{nr115} & Incorporating multivariate semantic association graphs into multimodal networks for information extraction from documents \\ \hline
116 & \cite{nr116} \citeyear{nr116} & Embedding Layout in Text for Document Understanding Using Large Language Models \\ \hline
117 & \cite{nr117} \citeyear{nr117} & LMDX: Language Model-based Document Information Extraction and Localization \\ \hline
118 & \cite{nr118} \citeyear{nr118} & GVDIE: A Zero-Shot Generative Information Extraction Method for Visual Documents Based on Large Language Models \\ \hline
119 & \cite{nr119} \citeyear{nr119} & LayoutPointer: A Spatial-Context Adaptive Pointer Network for Visual Information Extraction \\ \hline
120 & \cite{nr120} \citeyear{nr120} & One-Shot Transformer-Based Framework for Visually-Rich Document Understanding \\ \hline
121 & \cite{nr121} \citeyear{nr121} & Efficiency evaluation of filter sizes on graph convolutional neural networks for information extraction from receipts \\ \hline
122 & \cite{nr122} \citeyear{nr122} & GDP: Generic Document Pretraining to Improve Document Understanding \\ \hline
123 & \cite{nr123} \citeyear{nr123} & UNER: A Unified Prediction Head for Named Entity Recognition in Visually-rich Documents \\ \hline
124 & \cite{nr124} \citeyear{nr124} & OMNIPARSER: A Unified Framework for Text Spotting, Key Information Extraction and Table Recognition \\ \hline
125 & \cite{nr125} \citeyear{nr125} & RobustLayoutLM: Leveraging Optimized Layout with Additional Modalities for Improved Document Understanding \\ \hline
126 & \cite{nr126} \citeyear{nr126} & DocLLM: A Layout-Aware Generative Language Model for Multimodal Document Understanding \\ \hline
127 & \cite{nr127} \citeyear{nr127} & LiLTv2: Language-substitutable Layout-Image Transformer for Visual Information Extraction \\ \hline
128 & \cite{nr128} \citeyear{nr128} & EntityLayout: Entity-Level Pre-training Language Model for Semantic Entity Recognition and Relation Extraction \\ \hline
129 & \cite{nr129} \citeyear{nr129} & Light-Weight Multi-modality Feature Fusion Network for Visually-Rich Document Understanding \\ \hline
130 & \cite{nr130} \citeyear{nr130} & Language, OCR, Form Independent (LOFI) pipeline for Industrial Document Information Extraction \\
\bottomrule
\end{longtable}
\newpage
\section{Overview}

The individual papers were examined according to the following general properties, each of which considers different aspects, also with practical applications in mind.

\begin{itemize}
    \item Category: Indicates the superordinate method (see section 2.3), or a combination thereof, used specifically for the KIE task, if applicable. 
    
    \item Input modalities: Indicates, which modalities derived from input documents were considered for KIE. Besides textual, visual and layout-oriented features, we also include custom hand-crafted features in the analysis. Note that in case of LLM-based approaches, we do not consider the textual prompts to the LLM (required by design) as a text-based modality.

    \item Data basis: We gather the amount as well as types of documents used to implement and evaluate the proposed approaches, which provides an assessment of the underlying data basis and data requirements.

    \item Reproducibility \& deployability: These aspects highlight the availability of the implemented artifacts in terms of implemented code and/or model weights and whether they allow to be used in commercial settings.

    \item Independent of external OCR: Indicates whether the KIE approach requires no external OCR engine.

    \item Number of parameters: If specified by the authors, indicates the total number of model parameters (in millions) and thus the overall complexity of the model. If multiple model variants are proposed, the largest ones are listed.

    \item Integration of domain knowledge: Indicates whether the authors integrate domain knowledge (e.g., knowledge about business processes or values to be extracted) in some form.

    \item Evolution of existing KIE approach: Indicates whether the proposed approach is a successor to an already existing KIE method with distinct refinements.
    
    \item Evaluation in industry setting: Indicates whether the proposed approach has been evaluated in real-world industry settings - either in a quantitatively or qualitatively manner. This does not include cases where authors simply report the runtime of the approaches.
    
    \item Weakly annotated: Makes statements about the annotations required for the approach, i.e., whether fully-annotated documents including word-level annotations and bounding boxes are required or not. This therefore gives an indication of the effort required to train the corresponding approaches.

    \item Generative: Indicates whether the KIE system is an autoregressive approach that can produce arbitrary output during decoding steps.
\end{itemize}

\newpage

\tiny
\begin{longtable}{c|c|c|c|c|c|c|l|c|c|c|c|c|c|c|c|c|c}
\caption{Overview of the analysis} \label{overviewtab} \\
\hline
\multirow{2}{*}{\textbf{\#}} & \multirow{2}{*}{\textbf{Category}} & \multicolumn{4}{c|}{\textbf{Input modalities}} & \multicolumn{2}{c|}{\textbf{Data basis}} & \multicolumn{3}{c|}{\textbf{\makecell{Reproducibility \& \\ deployability}}} & \multirow{2}{*}{\rotatebox[origin=c]{90}{\textbf{Independent of external OCR}}} & \multirow{2}{*}{\rotatebox[origin=c]{90}{\textbf{Number of parameters (in M)}}} & \multirow{2}{*}{\rotatebox[origin=c]{90}{\textbf{Integration of domain knowledge}}} & \multirow{2}{*}{\rotatebox[origin=c]{90}{\textbf{Evolution of existing KIE approach}}} & \multirow{2}{*}{\rotatebox[origin=c]{90}{\textbf{Evaluation in industry setting}}} & \multirow{2}{*}{\rotatebox[origin=c]{90}{\textbf{Weakly annotated}}} & \multirow{2}{*}{\rotatebox[origin=c]{90}{\textbf{Generative}}} \\
\cline{3-6} \cline{7-8} \cline{9-11}
 &  & \rotatebox[origin=c]{90}{text} & \rotatebox[origin=c]{90}{image} & \rotatebox[origin=c]{90}{layout} & \rotatebox[origin=c]{90}{hand-crafted} & \rotatebox[origin=c]{90}{\makecell{number of documents used for KIE  \\ implementation}} & {\makecell{Document types for \\ KIE implementation}} & \rotatebox[origin=c]{90}{Code available} & \rotatebox[origin=c]{90}{Model weights available} & \rotatebox[origin=c]{90}{Commercial use} &  &  &  &  &  &  \\
\hline
1 & sequence & \ding{51} & \ding{55} & \ding{51} & \ding{51} & 326471 & Invoices & \ding{55} & \ding{55} & \ding{55} & \ding{55} & n/a & \ding{51} & \ding{55} & \ding{55} & \ding{55} & \ding{55} \\ \hline
2 & grid & \ding{55} & \ding{55} & \ding{55} & \ding{51} & 12000 & Invoices & $\sim$ & \ding{55} & \ding{51} & \ding{55} & n/a & \ding{55} & \ding{55} & \ding{55} & \ding{55} & \ding{55} \\ \hline
3 & grid & \ding{51} & \ding{55} & \ding{55} & \ding{51} & 12000 & Invoices & $\sim$ & \ding{55} & \ding{51} & \ding{55} & n/a & \ding{55} & \ding{51} & \ding{55} & \ding{55} & \ding{55} \\
\hline
\end{longtable}

\normalsize
\newpage
\begin{landscape}
\section{Sequence-based approaches}

\tiny
\begin{longtable}{>{\centering\arraybackslash}p{0.1cm}|>{\centering\arraybackslash}p{0.8cm}|>{\centering\arraybackslash}p{0.5cm}|p{2.35cm}|p{3.7cm}|p{2cm}|p{1.9cm}|p{2.55cm}|p{2.4cm}}
\caption{Analysis of sequence-based approaches} \label{sequencetab} \\
\hline
\multirow{2}{*}\textbf{{\#}} & \multirow{2}{*}{\textbf{Granularity}} & \multicolumn{3}{c|}{\textbf{Pre-Training}} & \multirow{2}{*}{\textbf{Textual encoder}} & \multirow{2}{*}{\textbf{Visual encoder}} & \multirow{2}{*}{\textbf{Decoder}} & \multirow{2}{*}{\textbf{Special attention mechanism}} \\
\cline{3-5}
& & \rotatebox[origin=c]{90}{\makecell{\# documents used \\ for Pre-Training}} & Datasets & Tasks & & & & \\
\hline
1 & word & / & / & / & / & / & LSTM &  \\ \hline
5 & character & / & / & / & / & Inceptionv3 & LSTM & Entity-aware attention mechanism \\ \hline
17 & word + character & / & / & / & FastText, ELMo, BERT & / & BiLSTM-CRF &  \\ \hline
18 & word & / & / & / & BiLSTM & / & LSTM & Global attention \\ \hline
20 & word & / & / & / & RoBERTa & / & Sequence labeling layer &  \\
\hline
\end{longtable}
\end{landscape}

\normalsize
\newpage
\begin{landscape}

\section{Graph-based approaches}

\tiny
\begin{longtable}{>{\centering\arraybackslash}p{0.1cm}| p{2.6cm}| >{\centering\arraybackslash}p{0.8cm}| p{3.4cm}| >{\centering\arraybackslash}p{0.3cm}| >{\centering\arraybackslash}p{0.1cm}| >{\centering\arraybackslash}p{0.9cm}| p{2.7cm}| p{2.5cm}| >{\centering\arraybackslash}p{0.2cm}| p{2cm}}
\caption{Analysis of graph-based approaches} \label{graphtab} \\
\hline
\multirow{2}{*}{\textbf{\#}} & \multirow{2}{*}{\textbf{Graph creation}} & \multicolumn{4}{c|}{\textbf{Nodes}} & \multicolumn{2}{c|}{\textbf{Edges}} & \multicolumn{2}{c|}{\textbf{Graph propagation}} & \multirow{2}{*}{\textbf{Method of KIE}} \\
\cline{3-6} \cline{7-8} \cline{9-10}
& & Granularity & Features & \rotatebox[origin=c]{90}{Feature vector size} & \rotatebox[origin=c]{90}{Global node} & Direction & Features & Propagation type & \rotatebox[origin=c]{90}{Layers} & \\
\hline
6 & Fully connected & segment & BiLSTM-embeddings & n/a & \ding{55} & directed & Horizontal and vertical distance + aspect ratio + relative height + relative width & GCN on node-edge-node-triplets with MLP and self-attention & 2 & BiLSTM-CRF \\ \hline
7 & Nearest neighbors in 4 major directions & word & Boolean features + relative distance to nearest neighbors + Byte Pair Encoding & 317 & \ding{55} & undirected & / & GCN with fourier transformation & 5 & Softmax layer (node classification) \\ \hline
9 & Nearest neighbors in 4 major directions & sentence & BiLSTM-embeddings + CharCNN-embeddings & 96 & \ding{55} & directed or undirected & Type (left-to-right, right-to-left, up-to-down, down-to-up) & GCN & 1 & BiLSTM-CRF \\ \hline
10 & k-NN (k=10) \& fully connected & word & Fasttext-embeddings + normalized coordinates & n/a & \ding{55} & undirected & / & GAT & n/a & MLP \\ \hline
11 & Landmark and field connected + pairwise connection between fields (36 rays emitted) & word & / & / & \ding{55} & directed & Normalized coordinates & MLP + CRF & / & Belief Propagation \\
\hline
\end{longtable}
\end{landscape}

\normalsize
\newpage
\section{Grid-based approaches}

\tiny
\begin{longtable}{>{\centering\arraybackslash}p{0.1cm}| >{\centering\arraybackslash}p{0.7cm}| >{\centering\arraybackslash}p{1.2cm}| >{\centering\arraybackslash}p{1.2cm}| >{\centering\arraybackslash}p{0.6cm}| p{2.9cm}| >{\centering\arraybackslash}p{1.1cm}| p{4.4cm}}
\caption{Analysis of grid-based approaches} \label{gridtab} \\
\hline
\multirow{2}{*}{\textbf{\#}} & \multirow{2}{*}{\textbf{Granularity}} & \multicolumn{3}{c|}{\textbf{Grid size}} & \multicolumn{2}{c|}{\textbf{Grid features}} & \multirow{2}{*}{\textbf{Method of KIE}} \\
\cline{3-5} \cline{6-7}
& & Height & Width & \rotatebox[origin=c]{90}{Feature vector size} & Content elements & Background elements & \\
\hline
2 & character & Image height & Image width & 54 & 1-hot encoded characters & all-zero vector & Semantic Segmentation, Bounding Box Regression (VGG-type) \\ \hline
3 & word & Image height & Image width & 768 & BERT-embeddings & all-zero vector & Semantic Segmentation, Bounding Box Regression (VGG-type) \\ \hline
4 & word & 512 & 512 & 35 & Character-based encoding of word + 1-hot encoded neighbor + RGB-channels & all-zero vector & Softmax-regression \\ \hline
8 & word & Image height & Image width & 4x128x103 & 1-hot encoded characters of N-grams & discarded & Custom (Attend to memory bank, Copy out attended information, parse into output) \\ \hline
12 & character & Image height & Image width & 257 & 1-hot encoded indexes of characters & all-zero vector & Semantic Segmentation (Coupled U-Net) \\
\hline
\end{longtable}

\normalsize
\newpage
\begin{landscape}

\section{LLM-based approaches}

\tiny
\begin{longtable}{>{\centering\arraybackslash}p{0.1cm}| p{1.3cm}| >{\centering\arraybackslash}p{0.1cm}| p{1.3cm}| >{\centering\arraybackslash}p{1.3cm}| >{\centering\arraybackslash}p{0.5cm}| p{1.7cm}| p{3cm}| >{\centering\arraybackslash}p{0.4cm}| p{1.7cm}| p{3cm}| >{\centering\arraybackslash}p{0.3cm}| >{\centering\arraybackslash}p{0.2cm}}
\caption{Analysis of LLM-based approaches} \label{llmtab} \\
\hline
\multirow{2}{*}{\textbf{\#}} & \multicolumn{2}{c|}{\textbf{Backbone}} & \multicolumn{2}{c|}{\textbf{Image encoding}} & \multicolumn{3}{c|}{\textbf{Pre-Training}} & \multicolumn{5}{c}{\textbf{KIE Fine-Tuning}} \\
\cline{2-3} \cline{4-5} \cline{6-8} \cline{9-13}
& LLM(s) & \rotatebox[origin=c]{90}{Open source/weights} & Encoder & Resolution & \rotatebox[origin=c]{90}{\# instructions} & Datasets & Tasks & \rotatebox[origin=c]{90}{\# instructions} & Datasets & Tasks & PEFT & \rotatebox[origin=c]{90}{\# trained parameters (M)} \\ 
\hline
80 & Davinci-003, ChatGPT & \ding{55} & / & / & / & / & / & / & / & / & \ding{55} & / \\ \hline
92 & mPLUG-Owl & \ding{51} & mPLUG-Owl & 224x224 (20 crops) & / & / & / & n/a & DeepForm, Kleister-Charity & Unified downstream task (VQA, IE, NLI, Image Captioning), Text Reading, Key Points Generation & LoRA & 86 \\ \hline
101 & Vicuna-7B & \ding{51} & Swin Transformer & 2560×2560 & 1520000 & Common Crawl*, Custom & Text detection, text recognition, text spotting, paragraph reading, full-text reading, image captioning & 370000 & FUNSD, SORIE, POIE & Text detection, text recognition, text spotting, paragraph reading, full-text reading, image captioning & \ding{55} & 7000 \\ \hline
102 & Llama-7B & \ding{51} & LayoutLMv3 & n/a & / & / & / & n/a & CORD, FUNSD & Summarization, Document QA & \ding{55} & 7000 \\
\hline
\end{longtable}
\end{landscape}

\normalsize
\newpage
\section{Evaluation}

\footnotesize
\begin{longtable}{>{\centering\arraybackslash}p{0.2cm}| >{\centering\arraybackslash}p{0.2cm}| >{\centering\arraybackslash}p{0.2cm}| >{\centering\arraybackslash}p{0.2cm}| >{\centering\arraybackslash}p{0.2cm}| >{\centering\arraybackslash}p{0.2cm}| >{\centering\arraybackslash}p{0.2cm}| p{2.3cm}| p{3.9cm}| >{\centering\arraybackslash}p{0.7cm}| >{\centering\arraybackslash}p{0.7cm}| >{\centering\arraybackslash}p{0.7cm}}
\caption{Analysis of evaluation methodologies} \label{evaltab} \\
\hline
\multirow{2}{*}{\textbf{\#}} & \multicolumn{2}{c|}{\textbf{\makecell{Evaluation \\ setup}}} & \multicolumn{2}{c|}{\textbf{Granularity}} & \multirow{2}{*}{\rotatebox[origin=c]{90}{\textbf{\textit{Unknowns} performance presented}}} & \multirow{2}{*}{\rotatebox[origin=c]{90}{\textbf{Benchmarks against related work}}} & \multirow{2}{*}{\textbf{Metrics}} & \multirow{2}{*}{\textbf{KIE datasets}} & \multicolumn{3}{c}{\textbf{F1 scores}} \\
\cline{2-3} \cline{4-5} \cline{10-12}
& \rotatebox[origin=c]{90}{Element-based} & \rotatebox[origin=c]{90}{String-based} & \rotatebox[origin=c]{90}{Overall performance presented} & \rotatebox[origin=c]{90}{Field-level performance presented} & & & & & CORD & FUNSD & SROIE \\
\hline
1 & \ding{55} & \ding{51} & \ding{55} & \ding{51} & \ding{55} & \ding{55} & Precision, Recall, F1 & Custom &  &  & \\ \hline
2 & \ding{55} & \ding{51} & \ding{55} & \ding{51} & \ding{55} & \ding{55} & dist & Custom &  &  & \\ \hline
3 & \ding{55} & \ding{51} & \ding{51} & \ding{51} & \ding{55} & \ding{51} & dist & Custom &  &  & \\ \hline
4 & n/a & n/a & \ding{51} & \ding{55} & \ding{55} & \ding{51} & Accuracy & Custom &  &  & \\ \hline
5 & \ding{55} & \ding{51} & \ding{51} & \ding{55} & \ding{55} & \ding{51} & meA, meP, meR, meF & Ticket &  &  & \\
\hline
\end{longtable}

\normalsize

\end{document}